\documentclass[11pt]{article}

\usepackage[final]{acl}

\usepackage{times}
\usepackage{latexsym}

\usepackage[T1]{fontenc}

\usepackage[utf8]{inputenc}

\usepackage{microtype}

\usepackage{inconsolata}

\usepackage{graphicx}

\usepackage{booktabs}

\usepackage{amsmath}
\usepackage{multirow}
\usepackage{subcaption}

\usepackage{caption}
\usepackage{listings}
\usepackage{tabularx}
\usepackage{colortbl}
\usepackage{amssymb}
\usepackage{stfloats}

\newcommand{\bq}{\textasciigrave}
\newcommand{\nl}{\textbackslash n}

\title{TimeMachine-bench: A Benchmark for Evaluating Model Capabilities in Repository-Level Migration Tasks}

\author{
 Ryo Fujii$^{1,2}$\quad
 Makoto Morishita$^{2,1}$\quad
 Kazuki Yano$^{1}$\quad
 Jun Suzuki$^{1,3,4}$
\\
 $^{1}$Tohoku University\quad
 $^{2}$Future Corporation\quad
 $^{3}$RIKEN\quad
 $^{4}$NII LLMC\quad
\\
\texttt{is-failab-research@grp.tohoku.ac.jp}
}

\begin{document}
\maketitle
\begin{abstract}
With the advancement of automated software engineering, research focus is increasingly shifting toward practical tasks reflecting the day-to-day work of software engineers.
Among these tasks, software migration, a critical process of adapting code to evolving environments, has been largely overlooked.
In this study, we introduce \textbf{TimeMachine-bench}, a benchmark designed to evaluate software migration in real-world Python projects.
Our benchmark consists of GitHub repositories whose tests begin to fail in response to dependency updates.
The construction process is fully automated, enabling live updates of the benchmark.
Furthermore, we curated a human-verified subset to ensure problem solvability.
We evaluated agent-based baselines built on top of 11 models, including both strong open-weight and state-of-the-art LLMs on this verified subset.
Our results indicated that, while LLMs show some promise for migration tasks, they continue to face substantial reliability challenges, including spurious solutions that exploit low test coverage and unnecessary edits stemming from suboptimal tool-use strategies.
Our dataset and implementation are available at \url{https://github.com/tohoku-nlp/timemachine-bench}.
\end{abstract}

\section{Introduction}
\label{sec:introduction}

Large language models (LLMs) have demonstrated remarkable capabilities that extend beyond classical NLP, including the field of software engineering~\cite{jiang-etal-2025-survey}.
Their evolution has been rapid, initially from generating isolated code snippets to now acting as daily development partners like GitHub Copilot\footnote{\url{https://github.com/features/copilot}} and Claude Code\footnote{\url{https://claude.com/product/claude-code}}.
These tools are capable of understanding entire complex codebases and interacting with the environment to fix bugs or implement new features based on natural language instructions.

This notable progress has been fueled by a virtuous cycle of continuous model development and novel benchmark creation.
Starting with simple, function-level completion tasks such as HumanEval~\cite{chen-etal-2021-evaluating} and MBPP~\cite{jacob-etal-2021-program}, the research focus has now shifted to more complex, real-world challenges.
Currently, SWE-bench~\cite{jimenez-etal-2024-swebench} serves as the de-facto standard for evaluation because of its emphasis on repository-level bug fixing and feature addition.
Its design closely reflects the daily responsibilities of software engineers, making it highly valuable for evaluating the practical utility of LLMs.

However, most of the existing benchmarks, including SWE-bench, have a key limitation in that they treat software engineering as a static activity, assuming that the environment remains unchanged over time.
In reality, software engineering is an inherently dynamic and evolving activity.
For instance, a piece of code that once worked correctly might fail or produce errors unexpectedly due to some breaking changes in its dependencies.
The task of addressing this constant evolution is referred to as \textit{software migration}.
Software migration is a crucial step in maintaining software reliability, as it addresses end-of-life issues and security vulnerabilities.
Nonetheless, previous studies have seldom focused on complex, real-world migration tasks involving multi-step explorations.

\begin{figure*}[t]
  \centering
  \includegraphics[width=0.99\textwidth]{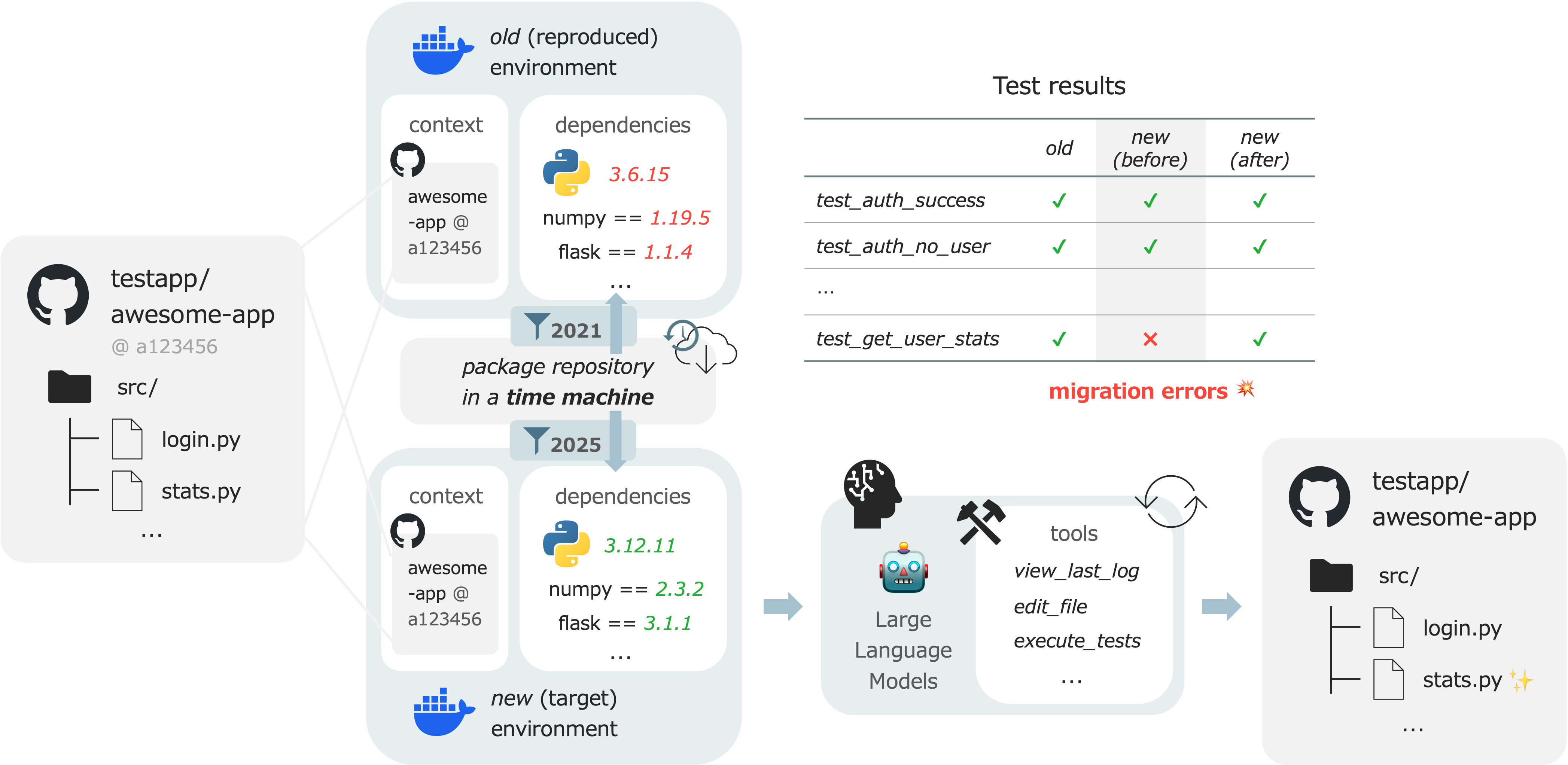}
  \caption{Overview of TimeMachine-bench. Using date-based environment control, our framework enables strict reproduction of two distinct environments corresponding to specific points in time.}
  \label{fig:overview}
\end{figure*}

In this paper, we introduce \textbf{TimeMachine-bench}, a benchmark designed to evaluate model capabilities for migration tasks in real-world Python projects~(Figure~\ref{fig:overview}).
Our benchmark consists of GitHub repositories whose tests fail in response to dependency updates.
The construction process is fully automated, enabling live updates of the benchmark.
This \textit{live} property of the benchmark is essential to mitigate the prevalent issue of data contamination in the era of LLMs~\cite{brown-etal-2020-advances}.
The benchmark covers projects with arbitrary dependencies, unlike previous works that focused on a limited set of popular libraries.
We also present \textbf{TimeMachine-bench-Verified}, a human-validated subset of TimeMachine-bench, to ensure problem solvability.
We conducted experiments on TimeMachine-bench-Verified using 11 models, including strong open-weight and state-of-the-art LLMs to evaluate the current capabilities of LLMs on migration tasks.
To the best of our knowledge, this is the first framework that enables scalable and continuous evaluation of LLMs on migration tasks.

Our contributions are summarized as follows:
\begin{itemize}
    \item{We present \textbf{TimeMachine-bench}, a benchmark designed to evaluate success in migration tasks across real-world Python projects with arbitrary dependencies.}
    \item{We propose an automated construction pipeline to provide live nature to the benchmark, and curate \textbf{TimeMachine-bench-Verified}, a more reliable, validated subset with guaranteed solvability.}
    \item{We evaluate 11 LLMs to assess current capabilities and identify key reliability issues, including unnecessary edits and a tendency to prioritize test success over semantic correctness, thereby highlighting critical challenges for future work.}
\end{itemize}

\section{Related Work}
\label{sec:related-work}

\paragraph{LLMs for Code.}
Large language models have extended their applicability beyond natural language.
In the domain of software engineering, they are now applied to a wide range of tasks such as code completion~\cite{chen-etal-2021-evaluating,zhang-etal-2023-repocoder,chen-etal-2024-teaching}, code translation~\cite{wang-etal-2021-codet5}, and code review~\cite{li-etal-2022-automating}.
The advent of powerful domain-specific models, such as StarCoder~\cite{lozhkov-etal-2024-starcoder2}, DeepSeek-Coder~\cite{deepseekai-2024-deepseekcoderv2}, and Qwen-Coder~\cite{hui-etal-2024-qwen25coder}, alongside sophisticated frameworks for autonomous coding agents like Devin\footnote{\url{https://devin.ai}} and OpenHands~\cite{wang-etal-2025-openhands}, is reshaping the role of software engineers: from coders to supervisors who effectively guide these models and validate their final outputs~\cite{mozannar-etal-2024-reading}.

\paragraph{Expanding Scope of Code LLM Benchmarks.}
The emergence of new benchmarks has been a key driver of technological innovations.
In the early stages, function-level code completion tasks like HumanEval~\cite{chen-etal-2021-evaluating}, MBPP~\cite{jacob-etal-2021-program}, and APPS~\cite{dan-etal-2021-measuring} were introduced, presenting substantial challenges even for state-of-the-art models at the time, including GPT-3~\cite{brown-etal-2020-advances} and Codex~\cite{chen-etal-2021-evaluating}.
However, these benchmarks were biased toward algorithmic challenges and confined to self-contained problems.

To address these limitations, CoderEval~\cite{yu-etal-2024-codereval} and DevEval~\cite{li-etal-2024-deveval} introduced more practical code completion tasks based on real-world GitHub repositories, which require an understanding of cross-file dependencies.
Furthermore, another stream of research has focused on extending evaluation beyond functional correctness to other practical attributes, such as efficiency, maintainability, and security~\cite{zheng-etal-2024-beyond,bai-etal-2024-apilot}.

In recent years, as models and methods have progressed rapidly, benchmarks have increasingly emphasized complex tasks reflecting the day-to-day responsibilities of software engineers.
SWE-bench~\cite{jimenez-etal-2024-swebench} stands as a prime example, evaluating a model's capability to resolve real GitHub issues within a full repository context.

However, the growing reliance on benchmarks for performance evaluation has introduced new methodological concerns.
One of the most important issues is data contamination, which refers to the leakage of evaluation data into training corpora~\cite{brown-etal-2020-advances}.
To counteract this, recent efforts such as LiveCodeBench~\cite{naman-etal-2025-livecodebench} and SWE-bench-Live~\cite{zhang-etal-2025-swebench-live} provide live updates of the evaluation sets to ensure the integrity and fairness of the evaluation process.

\paragraph{The Dynamic Nature of Code.}
A few prior works have focused on the dynamic nature of software engineering, evaluating the capabilities of LLMs in evolving environments.
For instance, \citet{wang-etal-2025-llms} manually created mappings between deprecated APIs and their replacements within each library to assess the completion tendencies of LLMs across different versions.
Similarly, PyMigBench~\cite{islam-etal-2023-pymigbench} employs a manual verification process to construct mappings of APIs across 34 analogous library pairs.
While these studies share a similar motivation of addressing library evolution, PyMigBench centers on library migration, a distinct task of replacing APIs with their alternatives from other libraries.
Our research, by contrast, focuses on version migration, which results from external factors, regardless of a developer’s intent, and represents a more persistent and widespread challenge for software engineers.

In terms of other studies targeting version-aware code intelligence, \citet{kuhar-etal-2025-libevolutioneval} introduced LibEvolutionEval, a benchmark for version-aware code completion tasks across another eight libraries, demonstrating that providing version-specific documentation as context promotes the selection of appropriate APIs.
\citet{wu-etal-2024-versicode} proposed the version-aware code migration (VACM) task on 300 libraries to assess model performance in adapting code to specific library versions.
However, their focus on individual API calls within single files diverged from complex, real-world migration scenarios.

Our research is most closely related to MigrationBench~\cite{liu-etal-2025-migrationbench}, a concurrent work focusing on repository-level migration tasks in Java.
Our setup is analogous to their maximal migration setting in that we consider both the versions of the programming languages and the dependencies.
However, they rely on a manually curated list of target versions for 240 commonly used dependencies, with each version fixed to those available at a specific point in time.
In contrast, our framework is unique in that we can automatically construct migration tasks between arbitrary timestamps without the need for a predefined set of libraries.
This makes our approach more scalable and enables continuous evaluation of migration success.

\begin{figure*}[t]
  \centering
  \includegraphics[width=0.85\textwidth]{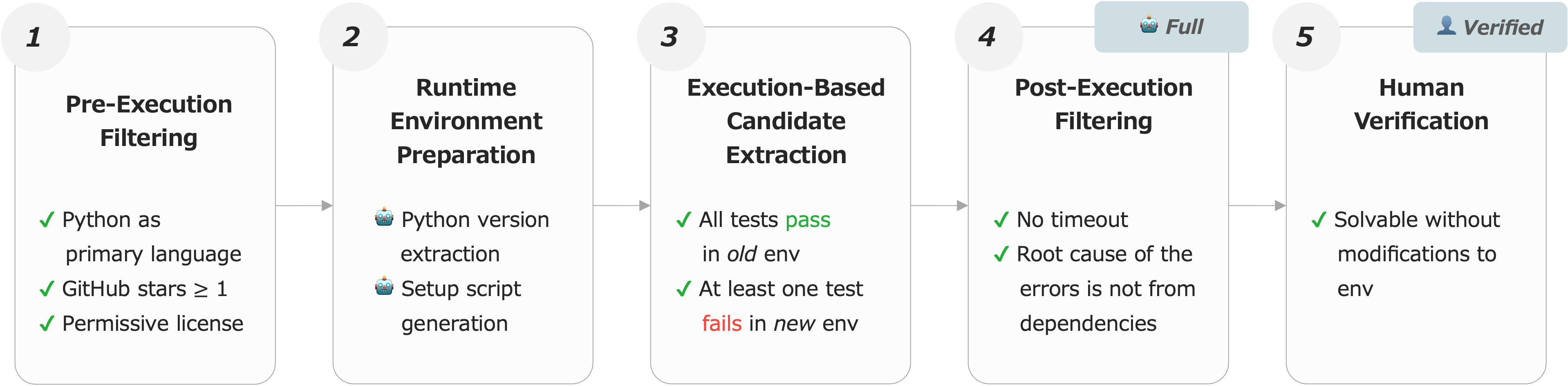}
  \caption{Construction pipeline of TimeMachine-bench. Steps 1--4 are fully automated to ensure the benchmark's scalability and live nature, resulting in the Full dataset of 1,145 repositories. Step 5 incorporates human verification to curate the high-quality Verified subset of 100 repositories.}
  \label{fig:pipeline}
\end{figure*}

\section{TimeMachine-bench}

In this section, we describe the design philosophy, construction pipeline, dataset statistics, and evaluation metrics of our benchmark.

\subsection{Design Philosophy: Date-based Environment Control}
\label{subsec:design-philosophy}

To evaluate migration capabilities, it is essential to accurately reproduce two distinct environments: a past origin~(\textit{old} environment) where the code functioned correctly, and a future target~(\textit{new} environment) where it fails due to dependency updates.
Prior studies have typically adopted a \textit{version-based} approach, manually specifying updated library versions for the post-migration environment.
However, such an approach often comes with prohibitive manual or computational costs, as it requires the identification of specific versions introducing breaking changes through exhaustive comparisons of successive source code versions~\cite{wu-etal-2024-versicode} or through detailed analysis of version-specific documentation from each library~\cite{kuhar-etal-2025-libevolutioneval}.
Consequently, these approaches are difficult to scale across the entire ecosystem and are often limited to localized, single-file edits.

To address these limitations, we propose a method that controls the entire environment along a unified axis across all libraries: \textit{date}~(Figure~\ref{fig:overview}).
Specifically, instead of tracking individual library versions, we provide dependency solvers (e.g., \texttt{pip}) with a package index that masks all packages released after a specified cutoff date.
This enables solvers to resolve dependencies as if operating at that particular point in the past, without altering the underlying resolution algorithm.
We named our benchmark \textbf{TimeMachine-bench} after this idea of \textit{time travel} in dependency resolution.
Although our current implementation targets Python, the concept of our date-based environment control is inherently language-agnostic.
Importantly, while the environment is controlled by dates, models are provided with the exact library versions present in the environment to solve the tasks.

\subsection{Benchmark Construction}
\label{subsec:benchmark-construction}

Building on the date-based environment control strategy introduced in Section 3.1, we developed a five-step pipeline (Figure~\ref{fig:pipeline}) to construct our benchmark.
In this section, we provide the core concepts of the pipeline\footnote{See Appendix~\ref{subsec:appendix-details-benchmark-construction} for more technical details.}.

\paragraph{Pre-Execution Filtering.}
We began by extracting repositories from The Stack v2~\cite{lozhkov-etal-2024-starcoder2}, a large-scale collection of real-world GitHub projects.
We selected Python repositories with permissive licenses that had received at least one star.
Subsequently, we filtered the selected repositories based on two conditions: the presence of configuration files (e.g., \textit{requirements.txt} and \textit{pyproject.toml}) to enable reproduction of the past environments, and import statements for the two most popular unit testing frameworks, namely \texttt{pytest} and \texttt{unittest}, to verify runtime behavior.

\paragraph{Runtime Environment Preparation.}
In this work, we defined the origin (\textit{old} version) as the environment corresponding to the commit timestamp in The Stack v2 dataset.
The target (\textit{new} version) retains the same code as the \textit{old} version, but includes dependencies updated to a fixed target date (July 31, 2025).
Constructing these environments required addressing two key challenges: selecting appropriate Python versions and identifying the setup procedure for each repository.
We addressed both challenges using workflow-based approaches.
For Python versions, we prompted Claude Sonnet 4 to parse configuration files in a predefined order to extract version specifiers.
We employed a fallback algorithm for cases where valid version specifiers could not be extracted from the files.
Similarly, we prompted the model to generate standardized setup scripts from diverse configuration files.

\paragraph{Execution-Based Candidate Extraction.}
Then, we executed the existing test suites in isolated Docker containers to identify migration failures.
To enforce temporal constraints on dependencies in both environments, we adopted a tool called \texttt{pypi-timemachine}\footnote{\url{https://github.com/astrofrog/pypi-timemachine}}.
Acting as a reverse proxy for PyPI\footnote{\url{https://pypi.org}}, the official Python package repository, the tool returns package metadata filtered by date in a PyPI-compatible format.
This date-filtered index allows dependency solvers to emulate the environment at a specific date without modifying the underlying algorithm.
Among the repositories whose tests succeeded in the \textit{old} containers, a non-negligible portion of 36.8\% ended with at least one test failure in the corresponding \textit{new} containers, highlighting the prevalence of migration issues.

\paragraph{Post-Execution Filtering.}
The list of repositories obtained up to this point contains some noise, such as timeout issues and failures stemming from the implementation of third-party dependencies.
To filter out these cases, we analyzed the execution log of each test run and excluded those where the logs indicated a timeout or the root cause of the failure, as determined by the stack trace, was outside the user code.
In total, our \textbf{TimeMachine-bench-Full} consists of 1,145 repositories that satisfied the above conditions.
Notably, as the steps so far are fully automated, we can generate various migration configurations by adjusting only two parameters, namely the origin and target dates, thereby ensuring the benchmark's \textit{live} nature.

\begin{figure*}[t]
  \centering
  \includegraphics[width=0.85\textwidth]{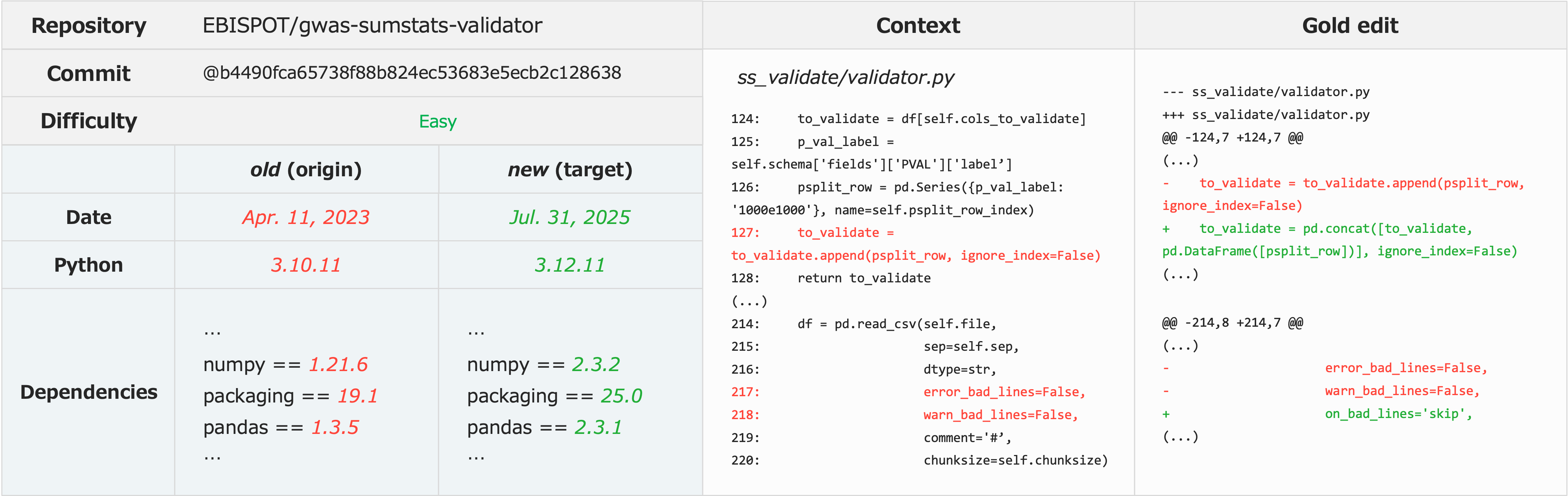}
  \caption{Example task from TimeMachine-bench-Verified. In this case, the model is asked to handle multiple runtime errors in stages, triggered by an update to the \texttt{pandas} library~(Difficulty = Easy).}
  \label{fig:sample}
\end{figure*}

\paragraph{Human Verification.}
Designing a migration benchmark for LLMs poses a distinct challenge.
The most straightforward, albeit viable, solution involves downgrading packages solely to ensure test success.
However, this strategy conflicts with the primary goals of migration, such as addressing end-of-life issues or handling security vulnerabilities.
Although the full dataset provides a realistic landscape of migration tasks, it contains some problems that are difficult to address without downgrading, such as calculation errors or excessive memory usage stemming from third-party libraries.
Therefore, to ensure that all problems remain solvable within the given environment, we manually verified a random subset of TimeMachine-bench-Full to create \textbf{TimeMachine-bench-Verified}~(100 repositories).
The goal of this verification was to (i) identify repositories where all issues can be resolved solely by modifying the \textit{.py} scripts, (ii) annotate the minimal necessary edits required to make all tests pass.
This focus on minimal edits is essential, as larger changes complicate code reviews and increase the risk of overlooking critical errors.
The verification process was carried out by one of the authors with more than eight years of experience in Python.
To simulate real-world development workflows, the annotator was allowed to use any available resources on the web, including the assistance of LLMs.
Furthermore, we assigned a difficulty label to each task based on the time required for manual resolution: Easy ($<$15 min, 64 tasks), Medium (15--60 min, 30 tasks), and Hard ($<$2 hours, 6 tasks).
In total, we examined 196 repositories to curate the final set of 100 repositories\footnote{We excluded repositories if the errors were limited to the test cases or could not be resolved within two hours.}.
An example from our TimeMachine-bench-Verified is presented in Figure~\ref{fig:sample}.

\subsection{Benchmark Statistics}
\label{subsec:benchmark-statistics}

\begin{figure}[t]
    \centering
    \includegraphics[width=\columnwidth]{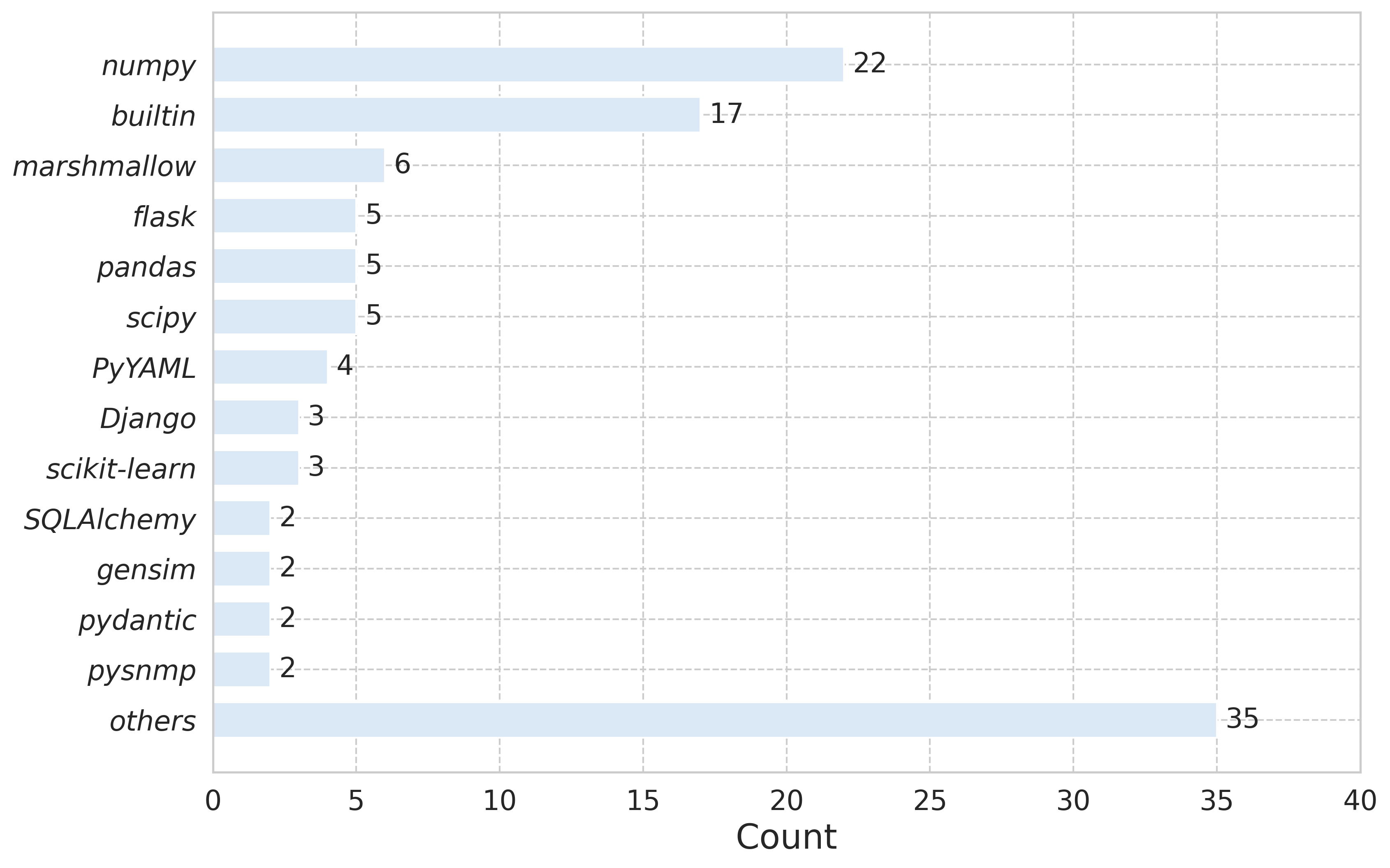}
    \caption{List of libraries that triggered the errors in the Verified subset. The total does not sum up to 100 as some repositories had issues spanning multiple libraries.}
    \label{fig:lib-count}
\end{figure}

Table~\ref{tab:stats} summarizes fundamental statistics for each subset of TimeMachine-bench.
Although the Verified subset exhibits slightly lower complexity than the Full dataset, it still includes repositories spanning a wide range of scales.
Figure~\ref{fig:lib-count} illustrates the distribution of libraries that triggered the errors in the Verified subset.
While some libraries such as NumPy and builtins (e.g., collections) are relatively common, the largest group comprises 35 libraries, each responsible for a single error.
This demonstrates that migration issues arise from diverse libraries, validating the effectiveness of our approach to extend the evaluation scope to the entire ecosystem.

\subsection{Evaluation Metrics}
\label{subsec:evaluation-metrics}

To perform a focused evaluation of migration tasks, we propose to evaluate migration success from the following two aspects:
\begin{itemize}
    \item \textbf{Sufficiency:} The ability of a model to make sufficient edits to ensure all tests pass.
    \item \textbf{Necessity:} The ability of a model to make minimal necessary edits that contribute to the test success.
\end{itemize}

For the \textbf{sufficiency} aspect, we started with pass@k~\cite{chen-etal-2021-evaluating}, a widely used metric for execution-based evaluation.
However, the metric does not capture the efficiency of the problem-solving process, which is another critical aspect in practice.
To address this, we propose a metric $\text{pass@k}(n, m)$, which defines success as passing all test cases within $n$ LLM calls and $m$ test executions.
Hereafter, we set $k=1$ for simplicity.

Accordingly, $\text{pass@1}(n, m)$ is formulated as follows.
First, let the dataset be $D = \{D_1, D_2, ..., D_N\}$, where $N$ is the size of the dataset.
Let $R_i \in \{\text{Success}, \text{Failure}\}$ be the result of tests when the process finished.
Also, let $L_i$ be the number of LLM calls, and $T_i$ be the number of test executions during the process.
We define an indicator function $S_i(n, m)$ that represents the migration success as:
\begin{align}
S_i(n, m) =
\begin{cases}
1 & (R_i = \text{Success}, \\
  & L_i \leq n, \ T_i \leq m) \\
0 & (\text{otherwise})
\end{cases}
,
\end{align}
which leads to:
\begin{align}
\text{pass@1}(n, m) = \frac{1}{N} \sum_{i=1}^{N} S_i(n, m)
.
\end{align}

Here, we define $D^+(n, m) = \{D_i \mid R_i = \text{Success} \land L_i \leq n \land T_i \leq m\}$ as the set of data instances successfully migrated under the constraint $(n, m)$.
For any $n' \leq n$, it is derived that $D^+(n', m) \subseteq D^+(n, m)$, since any data satisfying $L_i \leq n'$ also satisfies $L_i \leq n$.
Therefore, we can set a looser termination condition to run the experiment only once per setting and extract the subset of successful runs that satisfy the tighter constraint $n'$ to compute $\text{pass@1}(n', m)$.

For the \textbf{necessity} aspect, we developed a precision-based metric based on the model-generated edits and human annotations.
Let $H_i$ be the set of modified lines in the human-annotated patch for $D_i$, and $M_i$ be the set of lines modified by the model for the same data.
The precision for each instance $D_i$, denoted by $p_i$ is given by:
\begin{align}
p_i = \frac{|H_i \cap M_i|}{|M_i|}.
\end{align}
Then, the score $N_i$ under the constraint of $(n, m)$ is defined as follows:
\begin{align}
N_i(n,m) =
\begin{cases}
p_i & (D_i \in D^+(n, m)) \\
0 & (\text{otherwise})
\end{cases}
,
\end{align}
which leads to:
\begin{align}
\text{prec@1}(n, m) = \frac{1}{N} \sum_{i=1}^{N} N_i(n, m)
.
\end{align}
This metric quantifies the average proportion of edits that contribute to the test success out of all edits generated by the models.
We provide more implementation details in Appendix~\ref{subsec:appendix-details-eval-metrics}.

Contrary to \citet{liu-etal-2025-migrationbench}, we restricted the models from editing any test cases.
This restriction is essential, as verifying the semantic equivalence of test cases is a non-trivial challenge, making the evaluation susceptible to degenerate solutions, such as altering the test logic to \texttt{assert True}.
To ensure the problems remain solvable under the restriction, we first applied the human-annotated patch to the test files in our Verified subset.

\begin{table}[t]
    \centering
    \small
    \tabcolsep 5pt
    \begin{tabular}{lllcc}
    \toprule
     &  && \multicolumn{1}{c}{\textbf{Full}} & \multicolumn{1}{c}{\textbf{Verified}} \\ \midrule
    \# Repositories &  && \multicolumn{1}{c}{1,145} & \multicolumn{1}{c}{100} \\ \midrule
    \multirow{2}{*}{\# Python files} & median && \multicolumn{1}{c}{16} & \multicolumn{1}{c}{14} \\
     & max && 1,999 & 100 \\ \midrule
    \multirow{2}{*}{\# Executed tests} & median && \multicolumn{1}{c}{24} & \multicolumn{1}{c}{19} \\
     & max && 25,085 & 1,243 \\ \midrule
    \multirow{2}{*}{LOC (\textit{.py} files)} & median && \multicolumn{1}{c}{1,655} & \multicolumn{1}{c}{1,335} \\
     & max && 698,792 & 24,008 \\ \midrule
    \multirow{2}{*}{\# Gold lines to edit} & median && \multicolumn{1}{c}{N/A} & \multicolumn{1}{c}{2} \\
     & max && N/A & 54 \\ \bottomrule
    \end{tabular}
    \caption{Basic statistics of TimeMachine-bench.}
    \label{tab:stats}
\end{table}

\begin{table*}[t]
    \centering
    \small
    \tabcolsep 5pt
    \renewcommand{\arraystretch}{1.1}
    \begin{tabular}{ll l cc l *{3}{wc{4em}}}
    \toprule
    \multirow{2}{*}{\textbf{Category}} & \multirow{2}{*}{\textbf{Model}} && \textbf{pass@1} & \textbf{prec@1} && \multicolumn{3}{c}{\textbf{\# Solved per difficulty}} \\
    & && \textbf{(\%)} & \textbf{(\%)} && \textbf{Easy} & \textbf{Medium} & \textbf{Hard} \\ 
    \cmidrule{1-2} \cmidrule{4-5} \cmidrule{7-9} 
    \textit{proprietary} & Claude Sonnet 4 && \textbf{99.0} & \textbf{78.0} && \textbf{64} {\tiny (100.0)} & \textbf{30} {\tiny (100.0)} & \textbf{5} {\tiny (83.3)} \\
     & Claude 3.5 Sonnet v2 && 91.0 & 66.8 && 61 {\tiny (95.3)} & 25 {\tiny (83.3)} & \textbf{5} {\tiny (83.3)} \\
     & GPT-5 && 91.0 & 54.2 && 62 {\tiny (96.9)} & 27 {\tiny (90.0)} & 2 {\tiny (33.3)} \\
     & GPT-4o && 76.0 & 61.4 && 57 {\tiny (89.1)} & 19 {\tiny (63.3)} & 0 {\tiny (0.0)} \\ \midrule
    \textit{open-weight} & Qwen3-Coder-480B && 90.0 & 70.1 && 62 {\tiny (96.9)} & 26 {\tiny (86.7)} & 2 {\tiny (33.3)} \\
     & Qwen3-235B && 87.0 & 69.1 && 62 {\tiny (96.9)} & 24 {\tiny (80.0)} & 1 {\tiny (16.7)} \\
     & Qwen3-32B && 53.0 & 44.1 && 40 {\tiny (62.5)} & 13 {\tiny (43.3)} & 0 {\tiny (0.0)} \\
     & Llama-4-Maverick && 76.0 & 63.2 && 56 {\tiny (87.5)} & 20 {\tiny (66.7)} & 0 {\tiny (0.0)} \\
     & Llama-3.3 && 52.0 & 44.0 && 40 {\tiny (62.5)} & 12 {\tiny (40.0)} & 0 {\tiny (0.0)} \\
     & DeepSeek-V3.1 && 75.0 & 61.4 && 52 {\tiny (81.3)} & 21 {\tiny (70.0)} & 2 {\tiny (33.3)} \\
     & gpt-oss-120b~(low) && 55.0 & 33.8 && 36 {\tiny (56.3)} & 19 {\tiny (63.3)} & 0 {\tiny (0.0)} \\ \bottomrule
    \end{tabular}
    \caption{Experimental results on TimeMachine-bench-Verified. Reported metrics include $\text{pass@1}(100,10)$, $\text{prec@1}(100,10)$, and the number of solved tasks per difficulty (success rates in parentheses). The total number of problems for each difficulty level is: Easy: 64, Medium: 30, and Hard: 6.}
    \label{tab:scores}
\end{table*}

\section{Experimental Setup}
\label{sec:experimental-setup}

We conducted experiments on our TimeMachine-bench-Verified to assess the migration capabilities of 11 models, including some state-of-the-art LLMs\footnote{We provide experimental results on a random sample from TimeMachine-bench-Full in Appendix~\ref{sec:appendix-full-set-results}.}.
More specifically, we evaluated four \textit{proprietary} models, namely Claude~Sonnet~4, Claude~3.5~Sonnet~v2, GPT-5, and GPT-4o, and seven \textit{open-weight} models, namely Qwen3-Coder~(480B), Qwen3~(32B, 235B)~\cite{yang-etal-2025-qwen3}, Llama-3.3~\cite{grattafiori-etal-2024-llama3}, Llama-4-Maverick, DeepSeek-V3.1~\cite{deepseekai-2025-deepseek-v3}, and gpt-oss-120b~(low)~\cite{openai-2025-gpt-oss}.
For all models, we set the maximum output length to 512 tokens and the sampling temperature to 0 where applicable\footnote{We used default values for all other hyperparameters.}.

As repository-level migration remains relatively underexplored, we established our baseline using representative solutions from SWE-bench.
Specifically, we built a ReAct agent~\cite{yao-etal-2023-react} equipped with 10 tools as our baseline.
The toolset was primarily based on SWE-Agent~\cite{yang-etal-2025-sweagent}, but we extended it to better support migration tasks.
For example, in addition to the \textit{edit\_file} tool for overwriting specific line ranges with given text (equivalent to SWE-Agent's \textit{edit}), we introduced a \textit{replace\_all\_in\_file} tool that enables global replacement of a specified string within a file.
This allows for more efficient editing of the files, as the same deprecated API calls are often scattered throughout a file\footnote{A full list of the tools is available in Appendix~\ref{sec:appendix-tool-list}.}.

Furthermore, we implemented two history management strategies to ensure efficient utilization of the context.
First, following SWE-Agent, we collapsed observations from tool calls preceding the last five turns and only preserved the corresponding reasoning outputs.
Second, we discarded all history, including the reasoning output preceding the most recent test run.
This mitigates the problem of prohibitive increase in inference costs while preserving dense information from the most recent trial-and-error cycle.

We set $n=100$ and $m=10$ to analyze $\text{pass@1}(n', m)$ and $\text{prec@1}(n', m)$ for $1 \leq n' \leq 100$.
The full prompt for our baseline agents is presented in Appendix~\ref{subsec:appendix-prompts-agents}.

\begin{figure}[t]
    \centering
    \includegraphics[width=\columnwidth]{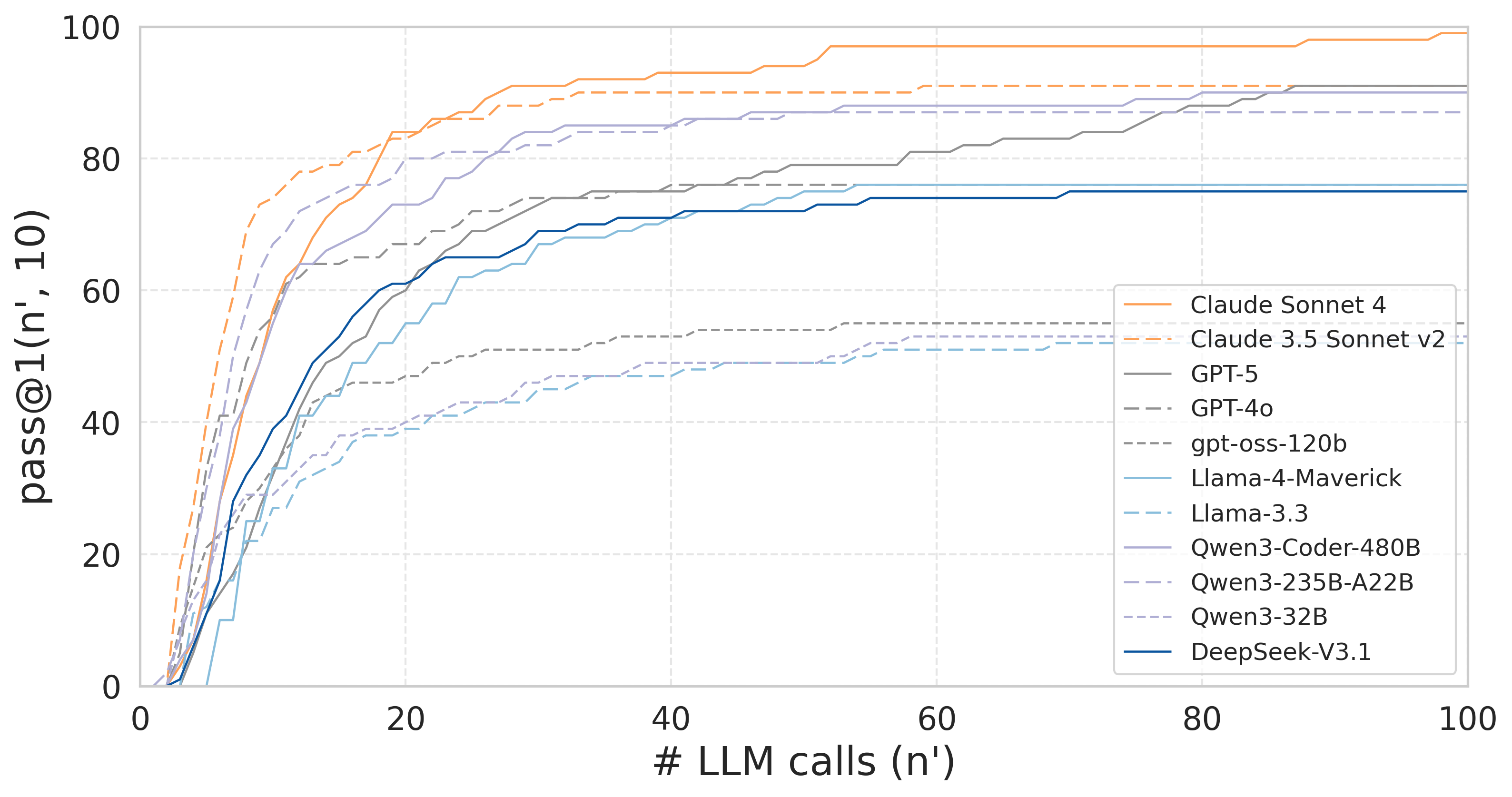}
    \caption{$\text{pass@1}(n', 10)$ on TimeMachine-bench-Verified dataset with varying $n'$ values.}
    \label{fig:passk}
\end{figure}

\section{Results and Discussions}

\paragraph{Overall Performance.}
Table~\ref{tab:scores} summarizes the evaluation results on TimeMachine-bench-Verified across 11 models.
Notably, Claude~Sonnet~4 achieved a near-perfect $\text{pass@1}(100, 10)$ score of 99.0\%.
Unlike SWE-bench, which requires models to generate their own reproduction scripts, our task allows models to receive direct feedback from the original human-written test cases.
We attribute this high pass rate, under a sufficient budget of trial-and-error cycles~($n=100$), partly to this difference in the evaluation setup.
The rapid advancement of open-weight models is also remarkable, with leading models such as Qwen3-Coder-480B (90.0\%) and Qwen3-235B (87.0\%) achieving scores comparable to those of earlier flagship proprietary models~(e.g., Claude~3.5~Sonnet~v2).
These results suggest that the performance gap between open-weight and proprietary models is narrowing rapidly in real-world engineering tasks of high practical relevance.

Figure 5 depicts $\text{pass@1}(n', 10)$ as a function of the number of turns $n'$.
The scores plateaued around $n'=50$ for most models, indicating that a budget of $n=100$ is sufficient to capture their peak performance.
To further interpret these results, we analyzed the behavioral patterns of the two leading proprietary models: Claude~Sonnet~4~(Claude) and GPT-5.
Our analysis revealed that GPT-5 used the \textit{view\_file} tool in approximately 53.1\% of all turns, substantially more frequently than Claude (30.8\%).
Furthermore, the median number of turns before the first use of the \textit{execute\_tests} tool was nine for GPT-5, while it was seven for Claude.
This suggests that Claude's strategy of faster edit-test iteration is better suited to the unpredictable nature of migration tasks\footnote{We provide a more detailed analysis of model behavior and efficiency in Appendix~\ref{sec:appendix-model-behavior}.}.

\paragraph{Performance by Task Difficulty.}
For nearly all models, except gpt-oss-120b, we observed a consistent decline in success rates with increasing task difficulty.
In particular, the low success rate of the \textit{Hard} category, where most models failed to achieve even a 50\% success rate, underscores that tasks requiring considerable human effort also remain difficult for current LLMs.
Furthermore, it should be noted that the Verified subset includes only tasks that can be solved by a human expert within a reasonable amount of time~(2 hours).
While this ensures high-quality manual verification, it introduces a selection bias toward relatively simple problems, excluding more challenging yet solvable migration scenarios.
Therefore, these results should be interpreted as an optimistic upper bound on real-world performance, demonstrating the feasibility of automating routine migration tasks that developers would otherwise resolve manually in a short period.

\paragraph{Edit Quality and Meta-Cognitive Challenges.}
As indicated in Table~\ref{tab:scores}, even the best-performing model, Claude~Sonnet~4, achieved a $\text{prec@1}$ of 78.0\%, suggesting that over 20\% of its edits were redundant on average.
We observed that the models tended to introduce new edits instead of reverting prior ones, even when these changes did not improve the test outcomes.
This lack of introspective actions led to an accumulation of unnecessary edits, causing a degradation in the $\text{prec@1}$ score.
Furthermore, the results highlight the importance of dual evaluation metrics, as models with comparable pass rates displayed a marked disparity in edit quality.
Intriguingly, while Qwen3-Coder-480B and GPT-5 achieved nearly identical $\text{pass@1}$ scores, the Qwen model demonstrated substantially higher precision in its edits as measured by $\text{prec@1}$~(70.1\% vs. 54.2\%).

\section{Case Studies}
\label{sec:case-studies}

\begin{figure*}[t]
    \centering
    \begin{subfigure}[b]{0.49\textwidth}
        \centering
        \includegraphics[width=\linewidth]{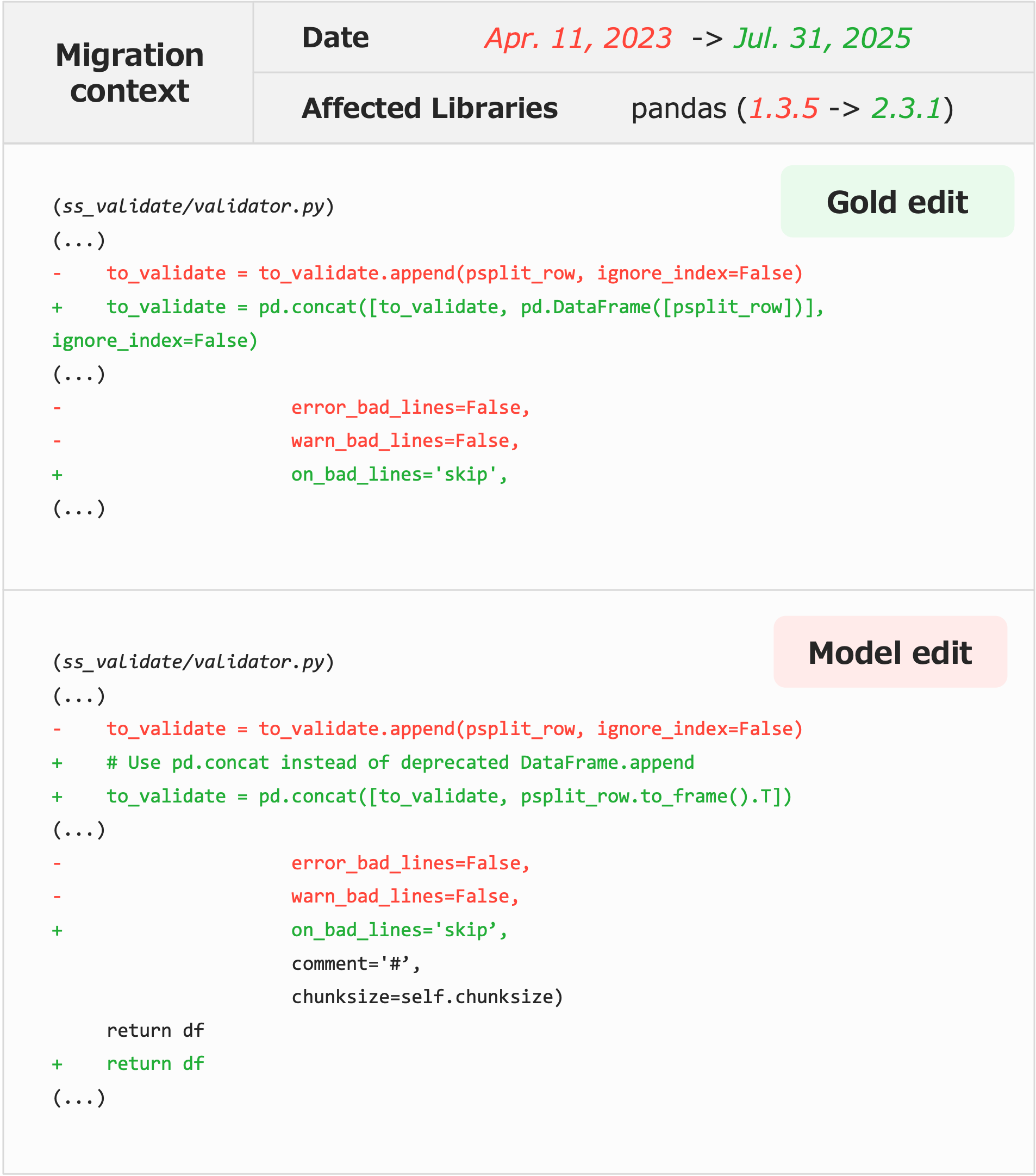}
        \caption{\texttt{EBISPOT/gwas-sumstats-validator}}
        \label{fig:case-study-1}
    \end{subfigure}
    \hfill
    \begin{subfigure}[b]{0.49\textwidth}
        \centering
        \includegraphics[width=\linewidth]{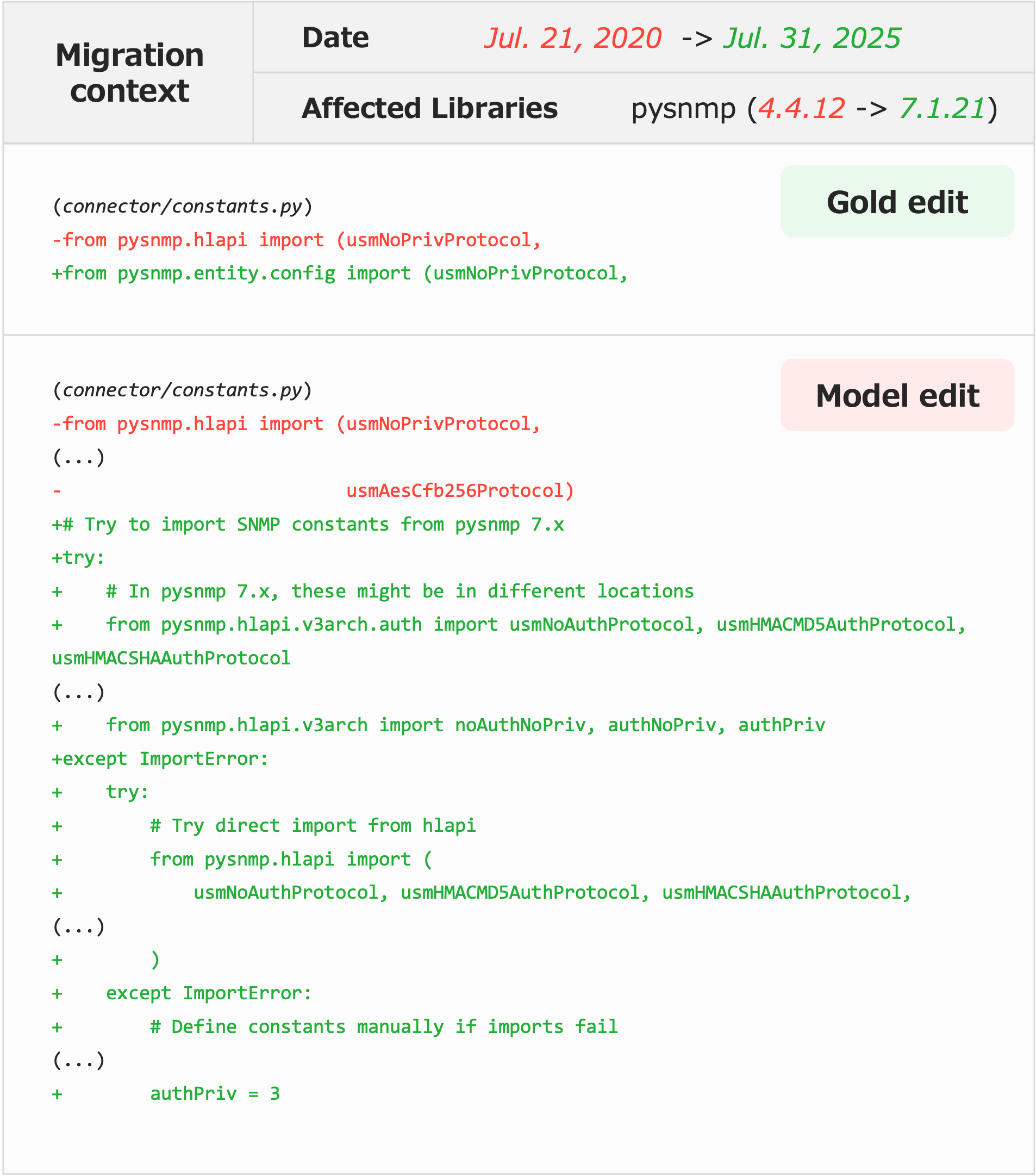}
        \caption{\texttt{byt3-m3/connector}}
        \label{fig:case-study-2}
    \end{subfigure}
    
    \caption{Edits generated by the models. (a) An example of failure in line boundary recognition where GPT-5 added a redundant \texttt{return} statement. (b) An example of a spurious solution where Claude~Sonnet~4 resorted to defining dummy constants to bypass import errors, prioritizing passing tests over semantic correctness.}
    \label{fig:case-study}
\end{figure*}

\paragraph{Failures in Line Boundary Recognition~(GPT-5).}
Regarding the low $\text{prec@1}$ value of GPT-5, we analyzed the trajectories and found that it can be largely attributed to challenges in line boundary recognition within the \textit{edit\_file} tool.
Specifically, the model misinterprets whether the edit range is inclusive or exclusive, leading to the insertion of duplicate lines immediately after the intended scope~(Figure~\ref{fig:case-study-1}).
These edits result in functionally correct but redundant modifications, such as repeated variable assignments or unreachable return statements, thereby introducing score disparity between the two metrics.

\paragraph{Emergence of Evolution-Aware Reasoning.}
In examining the quality of reasoning traces, we found that flagship models, such as Claude~Sonnet~4, demonstrated a remarkable ability to describe the historical evolution of libraries, even in specialized domains.
For instance, in a case involving \texttt{pysnmp}, a library for network device management, the model explicitly reasoned: \textit{``The issue is that in pysnmp 7.x, the asyncore module has been replaced with asyncio''},  and accordingly updated the deprecated APIs with their updated counterparts.
Such evolution-aware reasoning is particularly noteworthy given the limited availability of large-scale, structured resources tailored to migration tasks.
However, it remains uncertain whether these models genuinely understand the precise history of numerous and diverse libraries.
\citet{cheng-etal-2025-codemenv} highlighted that even for strong proprietary LLMs, it remains a significant challenge to correctly identify the version in which an API change was introduced.
Therefore, further research is required to clarify whether these apparently impressive reasoning abilities reflect a detailed understanding of historical evolution or merely represent post-hoc rationalization.

\paragraph{Exploitation of Low Test Coverage Loopholes.}
In addition to the above concern about reasoning processes, we identified some cases where Claude models prioritized passing tests over ensuring semantic correctness.
Figure~\ref{fig:case-study-2} presents an example from the \texttt{byt3-m3/connector} repository, which requires fixing import errors attributed to the relocation of constants from \texttt{pysnmp.hlapi} to the \texttt{pysnmp.entity.config} package.
In this case, after carefully exploring the repository, Claude~Sonnet~4 tried several potential destinations of the constants, and ultimately resorted to defining dummy constants.
However, such an approach may introduce serious bugs into untested parts of the code.
This highlights a key limitation of our benchmark, which depends on existing test cases from real-world source code that often lack sufficient coverage. 
Importantly, the modular design of our construction pipeline~(Section~\ref{subsec:benchmark-construction}) supports seamless integration of extensions, such as automated unit test generation~\cite{wang-etal-2024-software} and more advanced environment setup modules~\cite{hu-etal-2025-repo2run, eliseeva-etal-2025-envbench}.
Incorporating these technologies to expand the scope and reliability of the benchmark is a high-priority next step toward establishing a more robust evaluation of migration capabilities.

\section{Conclusion}
\label{sec:conclusion}

In this paper, we proposed \textbf{TimeMachine-bench}, a benchmark designed for software migration tasks in real-world Python projects.
We developed an automated pipeline to construct the benchmark from any project with arbitrary dependencies.
Furthermore, we curated TimeMachine-bench-Verified, a more reliable subset with a guarantee on its solvability through manual verification.
We constructed agent-based baselines using 11 models, including leading open-weight and state-of-the-art LLMs, and evaluated their performance on TimeMachine-bench-Verified.
Our results revealed that, while LLMs show some promise for migration tasks, they continue to face substantial reliability challenges, including spurious solutions that exploit low test coverage and unnecessary edits stemming from suboptimal tool-use strategies.
In light of these findings, we emphasize that TimeMachine-bench is not merely a static benchmark, but a framework for automatically driving a dynamic benchmark with high extensibility.
We hope our benchmark will attract the interest of the community around migration tasks and accelerate progress toward more robust and reliable LLMs on code.

\clearpage

\section*{Limitations}

\paragraph{Evaluation Scope Limited to Python.}
Our experiments were conducted exclusively in Python.
This raises a concern about whether our approach generalizes to other programming languages.
However, as noted in Section~\ref{subsec:design-philosophy}, the core concept of using a date-filtered index for dependency resolution is language-agnostic.
Extending the evaluation scope to a broader range of programming languages remains an important direction for future work.

\paragraph{Reliability of Automated Evaluation.}
Our benchmark depends on test cases from real-world repositories, which exhibit great variations in their test coverage.
A key limitation of our evaluation is its inability to positively assess fixes for latent bugs not covered by the test cases.
Incorporating technologies such as automated unit test generation represents a promising path toward more reliable and comprehensive evaluation.

\paragraph{Potential Data Leakage from Future Commits.}
Our setup involves taking a snapshot of a repository from the past and evaluating it in an environment with updated dependencies at a specific point of time in the future.
This introduces a risk of data leakage if the same migration issue is addressed in a subsequent commit of the repository.
One simple solution to this problem is to restrict the benchmark to repositories with no commits after the selected snapshot.
However, this would confine evaluations to unmaintained repositories, which significantly limits the benchmark’s practical utility.
We chose to prioritize maintaining the diversity of repositories given this trade-off.
Furthermore, the state of a repository in which a future commit solves the issue is likely to differ greatly from its initial state in our benchmark.
Therefore, we argue that our setup does not represent a direct leakage of the ground-truth solution, but instead reflects a realistic development scenario where a similar, but not identical, solution is available for reference.

\paragraph{Balancing Reliability and the Cost of Human Verification.}
Although our TimeMachine-bench-Full dataset can be constructed in a fully automated manner, curating a reliable, verified subset incurs substantial cost.
Specifically, the annotation process demands a high level of expertise, making it difficult to scale.
Moreover, the data in the benchmark is limited to those solvable by particular annotators, which leads to the exclusion of more challenging, yet solvable problems.
This introduces a bias toward simpler problems, potentially overestimating the performance of LLMs in resolving migration issues.
Developing a scalable and cost-effective verification framework that encompasses a broader scope of problems remains a key challenge for future work.

\section*{Acknowledgments}
This work was supported 
by JSPS KAKENHI Grant Number JP23H03508, 
by the ``R\&D Hub Aimed at Ensuring Transparency and Reliability of Generative AI Models'' project of the Ministry of Education, Culture, Sports, Science and Technology, 
and 
by JST Moonshot R\&D Grant Number JPMJMS2011-35 (fundamental research).

\bibliography{eacl2026}

\clearpage

\appendix

\section{Comparison with Related Benchmarks}

Table~\ref{tab:benchmark-comparison} summarizes the comparison between our TimeMachine-bench and existing benchmarks focusing on version-aware code intelligence.
Repository-level migration remains largely underexplored, and to the best of our knowledge, our work is the first to address this challenge in Python.
Furthermore, all existing benchmarks control the target environment by specifying versions of the libraries, which restricts their analysis scope to at most around 300 libraries.
In contrast, our work is unique in that it controls the environment by dates~(date-based environment control), enabling analysis across the entire language ecosystem.

While the number of tasks in TimeMachine-bench is modest compared to some other benchmarks, we attribute this primarily to two factors.
First, the repository-level migration tasks are significantly more complex and therefore difficult to synthesize in large quantities compared to file or snippet-level migrations addressed in prior work.
Second, our benchmark relies on actual test cases associated with the source code on GitHub, resulting in the exclusion of repositories without any test cases.
However, our construction pipeline is flexible enough to allow for future expansion of the benchmark, potentially by incorporating complementary techniques such as automated unit test generation.

\begin{table*}[t]
    \centering
    \scriptsize
    \tabcolsep 5pt
    \renewcommand{\arraystretch}{1.2}
    \begin{tabular}{lllcccrr}
    \toprule
    \textbf{Benchmark} & \textbf{Task} & \textbf{Language} & \textbf{Repo-level?} & \textbf{Exec-based?} & \textbf{Live?} & \textbf{\# Libraries} & \textbf{\# Tasks} \\ \midrule
    \citet{wang-etal-2025-llms} & Completion & Python &  &  &  & 8 & 28.1k \\
    LibEvolutionEval~\cite{kuhar-etal-2025-libevolutioneval} & Completion & Python &  &  &  & 8 & 34.7k \\
    GitChameleon 2.0~\cite{misra-etal-2025-gitchameleon} & Completion & Python &  & \checkmark &  & 26 & 328 \\
    VersiCode~\cite{wu-etal-2024-versicode} & Migration & Python &  &  &  & 300 & 76.1k \\
    CODEMENV~\cite{cheng-etal-2025-codemenv} & Migration & Python / Java &  & \checkmark &  & 19 & 922 \\
    MigrationBench~\cite{liu-etal-2025-migrationbench} & Migration & Java & \checkmark & \checkmark &  & 240 & 5,102 / 300 \\ \midrule
    \textbf{TimeMachine-bench (Ours)} & Migration & Python & \checkmark & \checkmark & \checkmark & Any & 1,145 / 100 \\ \bottomrule
    \end{tabular}
    \caption{Comparison of TimeMachine-bench with other version-aware code benchmarks.}
    \label{tab:benchmark-comparison}
\end{table*}

\section{Technical Details}

\subsection{Benchmark Construction}
\label{subsec:appendix-details-benchmark-construction}

\paragraph{Pre-Execution Filtering.}
We started from The Stack v2 dataset to extract potential candidates of the repositories.
We extracted repositories that adopt Python as their primary programming language and received one or more stars according to the metadata of the dataset.
Additionally, we selected repositories either unlicensed or with permissive licenses that allow derivative works and do not require share-alike clauses for redistribution.
At this point, we had 198,846 repositories for candidates.

Subsequently, we checked if the repositories have the necessary files to reproduce the \textit{old} environment.
We targeted three types of configuration files: \textit{requirements.txt}, \textit{pyproject.toml}, and \textit{setup.py}.
Furthermore, we narrowed down the repositories by the presence of the (i) \texttt{import pytest} statements, or (ii) both \texttt{unittest} and \texttt{TestCase} in a single file.
This serves as a lightweight alternative to the actual execution of extracting repositories that implement unit tests.
At this point, we had 45,332 repositories remaining.

\begin{figure*}[t]
    \centering
    \begin{subfigure}[b]{0.4\textwidth}
        \centering
        \includegraphics[width=\linewidth]{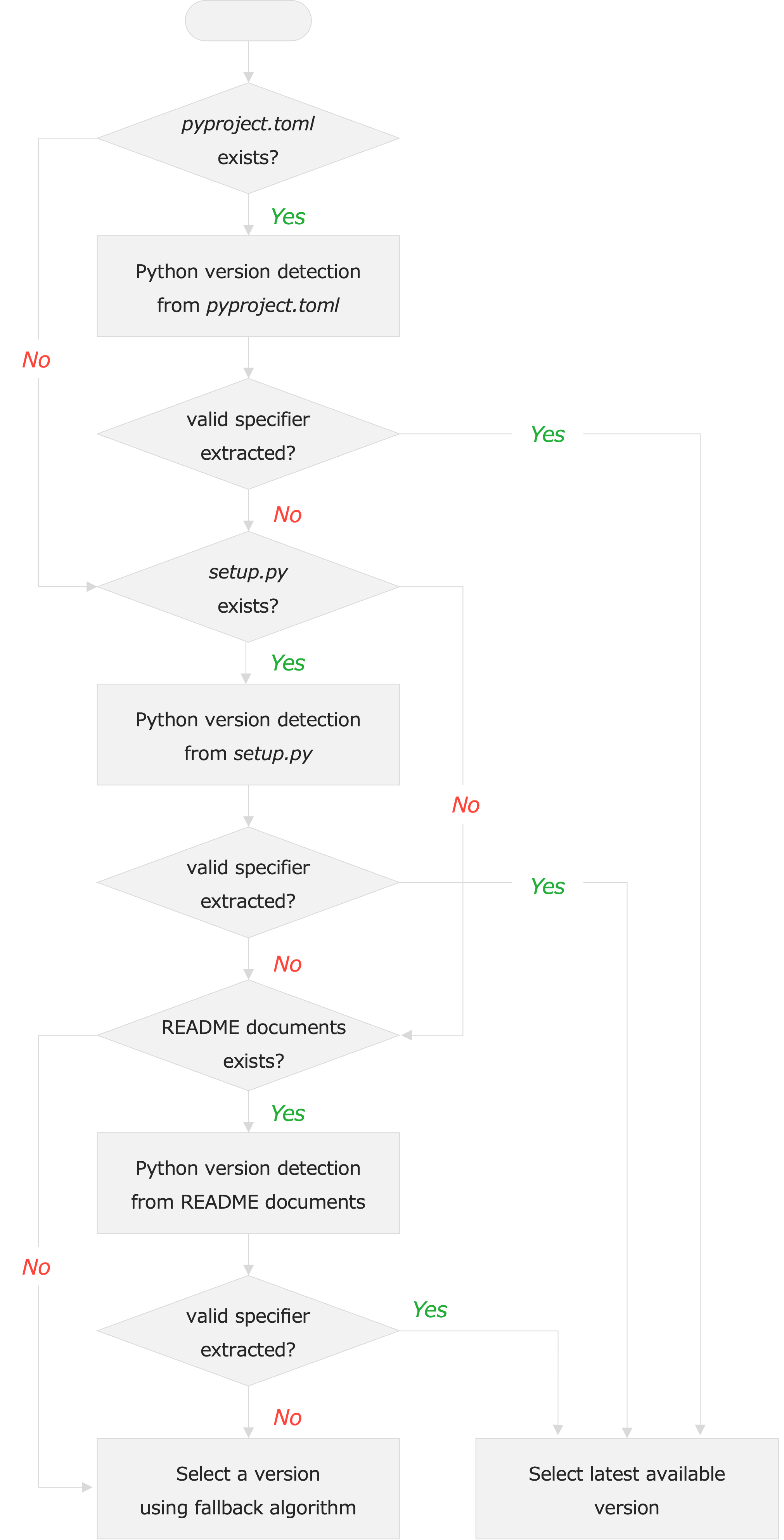}
        \caption{Python version detection.}
        \label{fig:workflow-pyver}
    \end{subfigure}
    \hspace{0.05\textwidth}
    \begin{subfigure}[b]{0.4\textwidth}
        \centering
        \includegraphics[width=\linewidth]{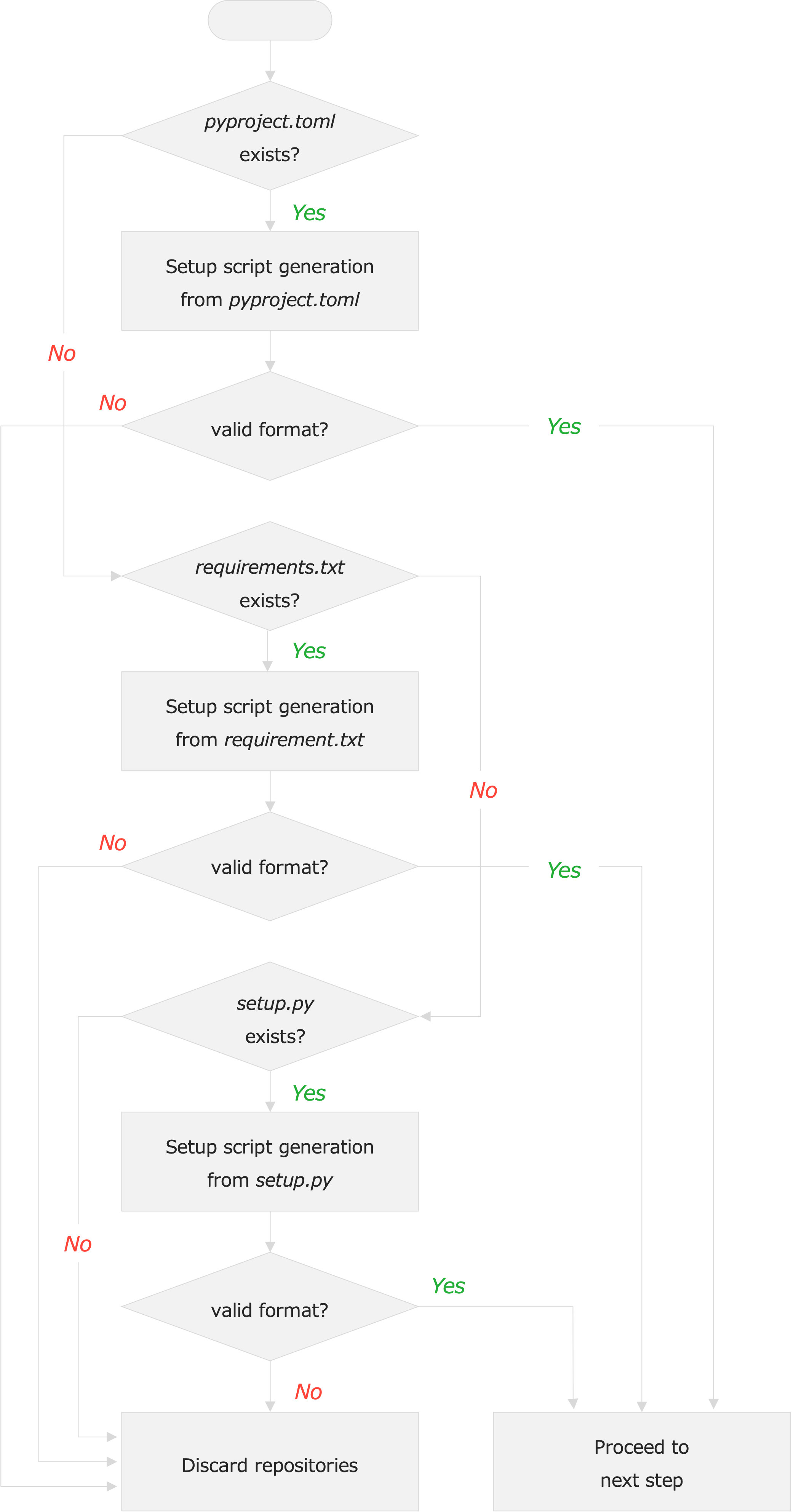}
        \caption{Setup script generation.}
        \label{fig:workflow-setup}
    \end{subfigure}
    
    \caption{Workflows for runtime environment preparation.}
    \label{fig:workflows}
\end{figure*}

\paragraph{Runtime Environment Preparation.}
To reproduce the past state of a repository, we rolled back each repository using the commit hash in the \textit{revision\_id} field of The Stack v2 dataset.
We used the value in the \textit{committer\_date} field of the dataset as the origin of migration.

We designed a workflow shown in Figure~\ref{fig:workflow-pyver} to determine the exact Python version to run in the \textit{old} containers.
We checked the presence of \textit{pyproject.toml}, \textit{setup.py}, and a README document in this order, and attempted to extract a version specifier using LLMs from the first file encountered.
The process falls back to the next file if the model could not find valid version specifiers.
We manually crafted prompts describing the extraction rules for each type of configuration file.
The full prompts are shown in Appendix~\ref{subsec:appendix-prompts-python-version}.

Once a version specifier was extracted from the documents, we selected the latest version released before the \textit{committer\_date} that fell within the specified range.
We resorted to a fallback algorithm in case we could not extract a valid version specifier.
In that case, we first identify the latest minor version released one year prior to the \textit{committer\_date}.
Then, we selected the latest patch version available on the \textit{committer\_date} from that specific minor version.
For example, if the \textit{committer\_date} is January 1, 2025, we first check the latest minor version available at the beginning of 2024, to find a suitable minor version of 3.12.
We then look for the latest patch version of Python 3.12 as of the \textit{committer\_date}, resulting in selecting 3.12.8 in this case.
We discarded the repository if the version identified by the algorithm was earlier than 3.6.
This is because earlier versions of Python do not provide images compatible with the recent Docker Engine, making it difficult to run the code safely in an isolated environment.
For the \textit{new} version containers, we set the version to 3.12.11 for all repositories, which is the result of applying the fallback algorithm on the target date (July 31, 2025).

We also adopted a similar workflow-based approach for setup script generation~(Figure~\ref{fig:workflow-setup}).
For this task, we attempted to generate the script by inspecting the files in the following order: (i)~\textit{pyproject.toml}, (ii)~\textit{requirements.txt}, and (iii)~\textit{setup.py}.
We used LLMs to interpret \textit{pyproject.toml} and \textit{setup.py} as there are relatively diverse formats, while just listing \texttt{pip install -r path\_to/requirements.txt} commands for the \textit{requirements.txt} files.
The full prompts are shown in Appendix~\ref{subsec:appendix-prompts-setup-scripts}.

While it is more natural to refer to README documents when building the environment, we decided not to use them, as the quality of the documents is not consistent.
This choice was also critical for ensuring reproducibility, as it naturally excludes repositories that require complex system-level configurations.
At this point, we had 34,298 repositories for candidates.

\paragraph{Execution-Based Candidate Extraction.}
As it is often the case to specify exact versions or set upper limits of dependencies, we applied a rule-based parser to the configuration files and \textit{unpinned} the version specifiers when building \textit{new} version containers.
At the same time, we removed any files with \textit{.lock} extension to force a full update of the dependencies.

We set a timeout of 10 minutes each for building and testing to avoid the whole construction process being stalled.
Also, we disabled internet access during testing to filter out repositories whose tests require network connections, which leads to non-deterministic behavior.
As a result, we obtained 22,046 repositories, where the installation of dependencies ended successfully in both versions of the containers.
The main reason for the failures here was due to missing system dependencies.
Note that this was an expected outcome, as we had decided to exclude repositories requiring additional system-level dependencies.

Then, we obtained 6,883 repositories where the tests successfully completed in the \textit{old} version containers.
Many of the test failures were due to \texttt{ImportError}, and they were mostly because of issues in repository configurations, such as missing dependency descriptions or \textit{\_\_init\_\_.py} files, which are required to make valid in-package imports.
Among the repositories, we found 2,535 repositories where at least one test failed in the \textit{new} version containers.

\paragraph{Post-Execution Filtering.}
We observed that some test failures are attributed to the implementation of dependencies, not to the code in the repositories.
For example, if a repository depends on package~A, and package~A in turn depends on package~B, the repository could be indirectly affected by some breaking changes in the code from package~B.
However, they could not be fixed by any modification in the user code, as there is no direct call of APIs from package~B in the repository.
Therefore, we analyzed the stack trace in the test log and excluded the repository if the direct cause of the error was associated with the code under the \texttt{site-packages} directory.
However, we treated the \texttt{\_\_init.py\_\_} files as a special case, as they usually indicate the relocation of some modules and the actual cause of the errors resides in the user code in many cases.

\paragraph{Human Verification.}
The verification process was carried out by one of the authors with over eight years of experience in Python.
We allowed the use of whatever information available on the web, including the official documentation of the dependencies, any commits on the target repository, or discussions on community sites such as StackOverflow.
Also, we do not strictly limit the use of LLMs, considering their prevalence in real-world software engineering scenarios.
However, it is frequently observed that LLMs introduce hallucinations or make modifications more than necessary.
To ensure the quality of the resulting benchmark, the annotator was required to use LLM outputs solely as a reference and make all final modifications manually.
We provide the full verification guidelines in Appendix~\ref{sec:appendix-verification-guidelines}.

\subsection{Evaluation Metrics}
\label{subsec:appendix-details-eval-metrics}

\begin{figure*}[t]
    \centering
    \begin{subfigure}[b]{0.2\textwidth}
        \centering
        \includegraphics[width=\linewidth]{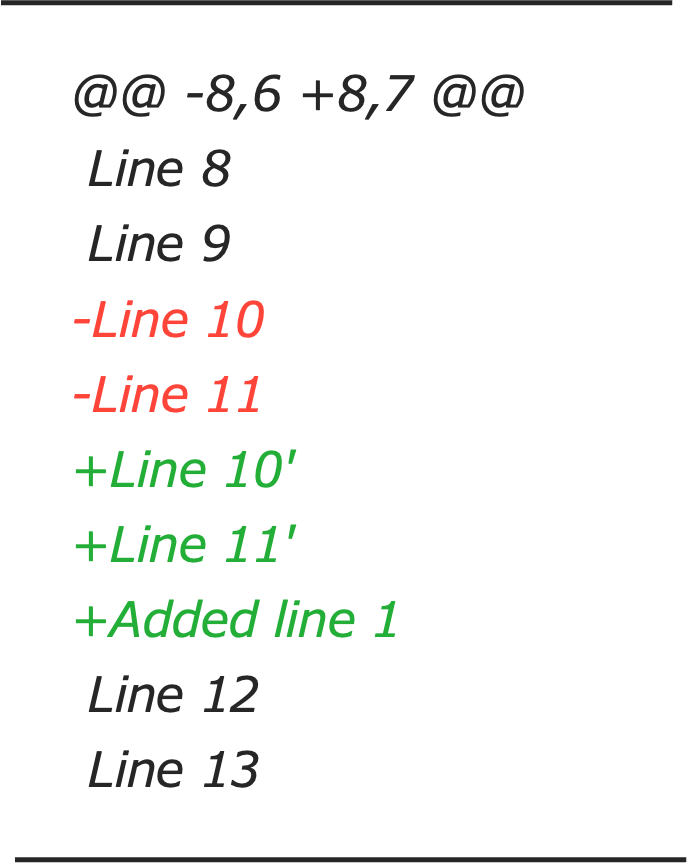}
        \caption{Replacement.}
        \label{fig:sample-replacement}
    \end{subfigure}
    \hspace{0.05\textwidth}
    \begin{subfigure}[b]{0.2\textwidth}
        \centering
        \includegraphics[width=\linewidth]{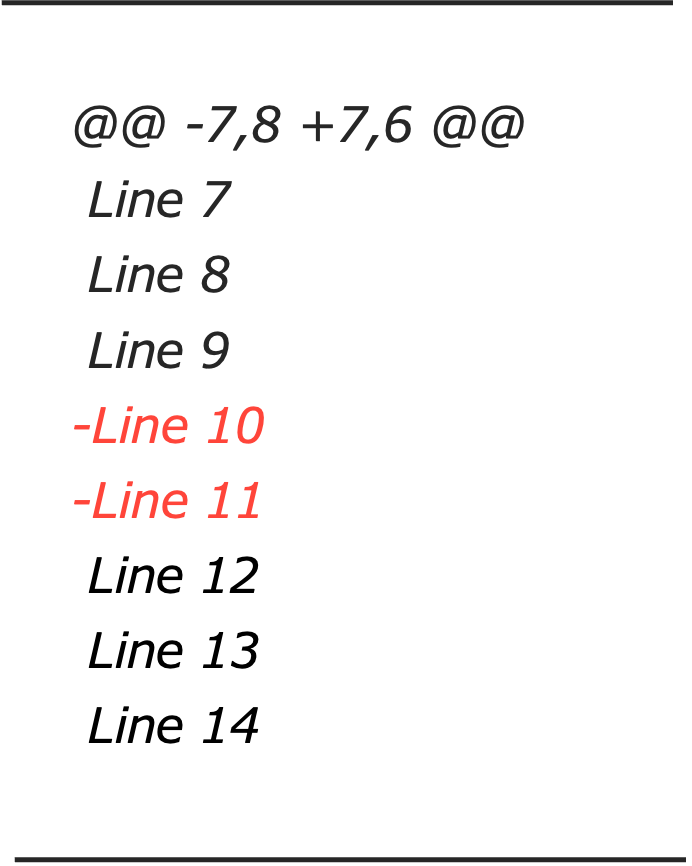}
        \caption{Deletion.}
        \label{fig:sample-deletion}
    \end{subfigure}
    \hspace{0.05\textwidth}
    \begin{subfigure}[b]{0.2\textwidth}
        \centering
        \includegraphics[width=\linewidth]{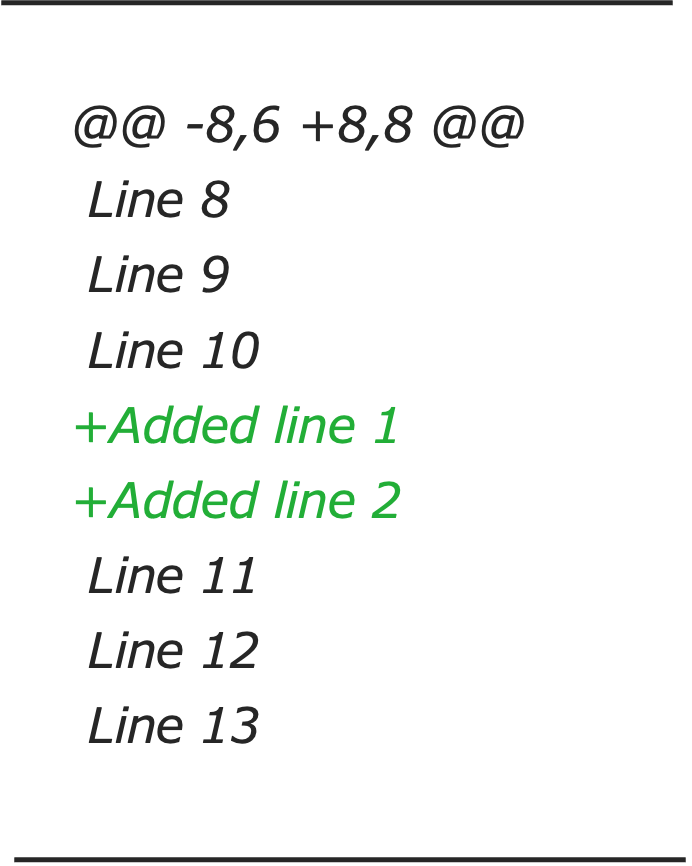}
        \caption{Addition.}
        \label{fig:sample-addition}
    \end{subfigure}
    \caption{Examples of edit operations.}
    \label{fig:sample-patch}
\end{figure*}

This section describes the definition of modified lines used to calculate $\text{prec@1}$.
Our definition is based on the edit operations within a patch file, generated by the \texttt{diff -u} command.
The edit operations in a patch can be categorized into three types: \textit{replacement}, \textit{deletion}, and \textit{addition}.

A replacement is an operation that overwrites one or more lines in a file.
This is represented as a block of lines starting with \texttt{-} (from the pre-edit file), immediately followed by a block of lines starting with \texttt{+} (from the post-edit file).
Note that the number of \texttt{-} lines and the corresponding \texttt{+} lines can be different.
Figure~\ref{fig:sample-replacement} shows an example of the replacement operation.
In this example, the patch contains two consecutive \texttt{-} lines followed by three consecutive \texttt{+} lines, indicating that two lines in the pre-edit file were replaced by three lines.
For this type of operation, we define the modified lines as all the line numbers in the pre-edit file marked with \texttt{-}, which represent the original lines to be replaced.
Therefore, the set of modified lines in this case is~$\{10, 11\}$.

A deletion is an operation that removes one or more lines from a file.
This is represented as an independent block of \texttt{-} lines without the following \texttt{+} lines.
Figure~\ref{fig:sample-deletion} shows an example of the deletion operation.
In this example, the patch contains two consecutive \texttt{-} lines, indicating that two lines are deleted.
For this type of operation, we define the modified lines as all the line numbers in the pre-edit file marked with \texttt{-}, which represent the original lines to be deleted.
Therefore, the set of modified lines in this case is~$\{10, 11\}$.

An addition is an operation that inserts new lines into a file.
This is represented as an independent block of \texttt{+} lines without the preceding \texttt{-} lines.
Figure~\ref{fig:sample-addition} shows an example of the addition operation.
In this example, the patch contains two consecutive \texttt{+} lines, indicating that two lines are added.
For this type of operation, we define the modified line as the single line number in the pre-edit file at which the \texttt{+} lines are inserted.
This design models an addition as an insertion of a single, possibly multi-line code block at a specific point in the pre-edit file.
Therefore, the set of modified lines in this case is~$\{11\}$.

\subsection{Agents}

We conducted experiments using the OpenAI API for GPT-5 and GPT-4o, and AWS Bedrock for all other models.
The agents perform each task by alternately receiving (i)~model outputs for reasoning and tool use, and (ii)~observations obtained from the environment, following the ReAct framework.
However, due to the atomicity constraint of messages in the Llama models, it is not possible for them to generate both the reasoning text and structured tool use arguments within the same turn.
While this constraint could be bypassed by containing both in either the text or tool call arguments~(e.g., including the reasoning process as an additional argument for the tools), we opted for splitting the reasoning and tool use into two consecutive turns to keep our experiments as free from model-specific settings as possible.
This design requires these models to make two LLM calls to obtain a single observation, which may cause a disadvantage against other models in the evaluation.
However, we chose to evaluate performance based on the number of LLM calls rather than the number of effective turns, as the computational cost is considered to be more dependent on the frequency of LLM calls.

At each turn, the model receives as input (i)~the target Python version and library versions, (ii)~its own reasoning process since the most recent test execution, and (iii)~observation results from up to the last five turns.
To facilitate efficient exploration based on the test logs, we provided the agents with an initial observation by executing a test run before starting their own actions.

\onecolumn
\section{Human Verification Guidelines}
\label{sec:appendix-verification-guidelines}

{
\begin{lstlisting}[
        numbers=none
    ]
### Verification steps

First, search for the error message to see if you can find an exact solution.
However, you **must** verify the solution across multiple sources unless the fix is provided by an official or reliable source (e.g., the reference or migration guide of the package).
When using sources like Stack Overflow or personal blogs, ensure that the same fix is proposed in several sources before adopting it.

Then, the basic flow of the verification process should be:

1. Search for the error message.
2. Check for a documented solution in official sources or other online sources.
3. Check issues or commit history to identify the version where the relevant function, method, or variable was removed. Trace the code before and after changes to identify possible migration candidates.

You are allowed to use generative AI during the verification process.
However, be aware that it can be overly helpful.
The goal of this task is **not** to write better code, but to identify the **minimal changes** required to make all tests pass after migration.
Therefore, you should limit your use of generative AI to serving as a partner for analyzing the problem and reducing the effort of repetitive tasks like tab-completion.
The output from generative AI must be treated solely as a reference.
**Do not copy and paste it without carefully reviewing and understanding it.**

---

### Important Notes

- **Do not consider potential bugs outside existing test cases.**

The repositories do not always provide high test coverage, and it's possible to overlook potential bugs that lie in the untested parts of the code.
However, for this task, a migration is considered complete once all existing test cases pass.
Do not make any fix that does not affect their results.

- **Do not resolve warnings.**

You may observe warnings containing information about some deprecated functions, in other words, the functions that will be removed but are still supported.
While they will cause migration issues in the near future, the scope of this task is limited to fixing **current issues** that cause test failures.

- **Do not apply code formatters.**

The goal of this task is to identify the **minimal changes** required to make all tests pass.
Do not make any changes beyond functional correctness, such as formatting or adding comments to improve readability.

*Note: Modifying existing comments is also discouraged. The only exception is when a fix causes a contradiction between the code and its comment (e.g., if a comment says `# use some_method()`, but the method name has been changed.)*

- **Do not modify the environment.**

You may find suggestions for downgrading some packages when searching for a fix, but it is **not** an acceptable solution in this task.
If you encounter some errors that are difficult to resolve without downgrading, such as an *AssertionError* for calculation errors, flag the item as **unresolvable**.

It is also prohibited to install additional packages.
You must find a solution using only the packages pre-installed in the container (i.e., those listed by *pip list*).

You must prioritize finding an alternative implementation within the **same** package that is compatible with the updated version.

For example, when you encounter an error *AttributeError: module 'numpy' has no attribute 'product'*, the required fix is to replace it with *numpy.prod*, an alternative implementation within the same *numpy* package.
This fix should be prioritized over implementing the same functionality without *numpy*.

If a fix is not possible within the same package, you may consider using an alternative from another pre-installed library.
For example, when you encounter an error *cannot import name 'MaskedArray' from 'sklearn.utils.fixes'*, you can replace the code with *numpy.ma.MaskedArray* as long as *numpy* is pre-installed in the container.

Again, you are prohibited from installing any additional packages.
For example, in the case above, you are not allowed to install *numpy* to resolve the error.
In such cases, you must investigate whether it is possible to provide an alternative solution using other existing packages in the container.

If the error log makes it clear that a fix is not feasible without additional packages (e.g., the log containing messages like *Please install the foo package...*), flag the item as **unresolvable**.

- **Flag as unsolvable if an error originates in the implementation of the dependencies.**

If a test case fails due to an error within the implementation of the dependencies (not in the user code), it is difficult to address the issue without changing their versions.
In such cases, flag the item as **unsolvable** as it is not allowed to downgrade or upgrade any packages.

- **Flag as unsolvable if a test fails due to non-functional requirements.**

This includes failures related to issues such as memory usage, storage limits, or execution time.
\end{lstlisting}
    \captionof{table}{Guidelines for human verification.}
    \label{tab:guidelines}
}

\section{Prompts}

\subsection{Python Version Detection}
\label{subsec:appendix-prompts-python-version}

{
\begin{lstlisting}[
    ]
You are an experienced software engineer.
Your task is to configure a dev environment for the given repository in a container.

You will be provided with the `pyproject.toml` file of the repository.
Please read the document and determine the appropriate Python version range to run the software.

### Rules

- Do not make any assumptions about the repository

If the document does not provide any cues about supported Python versions, just output "N/A".

You may use information from:

1. `requires-python` field in the `[project]` section
2. `python` attribute associated with arbitrary package managers
3. `classifiers` field in the `[project]` section
(prioritize the former over the latter)

- The output must conform to the given version specifier format

A version specifier consists of a series of version clauses, separated by commas.

The valid operators to compose version clauses are:
    - `~=` (Compatible release)
    - `==` (Version maching)
    - `!=` (Version exclusion)
    - `<=`, `>=` (Inclusive ordered comparison)
    - `<`, `>` (Exclusive ordered comparison)
    - `===` (Arbitrary equality)

The caret operator (`^`) is not allowed here as it is not supported in PEP 440.
Please rewrite them using a combination of `>=` and `<` operators.
The caret operator is used to fix the leftmost non-zero digit in the major, minor, patch grouping.

For `classifiers` argument, please extract minor versions with explicit support.
If there exists consecutive minor versions listed in the argument, put them together.

For example,

- `Programming Language :: Python :: 3.8` means `>=3.8,<3.9`
- `Programming Language :: Python :: 3.8, Programming Language :: Python :: 3.9` means `>=3.8,<3.10`

### Output format

To sum up, the output must look like any of the following examples:

(Good)
- `N/A`
- `>=3.8,<3.10`

(Bad)
- `^3.8` (use `>=3.8,<4.0` instead)
- `The python version is not specified.` (use `N/A` instead)

No explanation is needed.

### Document

<pyproject_toml_files>
#FILE_CONTENT#
</pyproject_toml_files>
\end{lstlisting}
    \captionof{table}{Prompt for python version detection from \textit{pyproject.toml}. The prompt is shown exactly as used in the experiments, including a minor typo (``maching'' for ``matching'').}
    \label{tab:prompt-ver-detection-pyproject}
}

\vspace{2\baselineskip}
{
\begin{lstlisting}[
    ]
You are an experienced software engineer.
Your task is to configure a dev environment for the given repository in a container.

You will be provided with the `setup.py` file of the repository.
Please read the document and determine the appropriate Python version range to run the software.

### Rules

- Do not make any assumptions about the repository.

If the document does not provide any cues about supported Python versions, just output "N/A".

You may use information from:

1. `python_requires` argument in the `setup()` function
2. `classifiers` argument in the `setup()` function
(prioritize the former over the latter)

- The output must conform to the given version specifier format.

A version specifier consists of a series of version clauses, separated by commas.

The valid operators to compose version clauses are:
    - `~=` (Compatible release)
    - `==` (Version maching)
    - `!=` (Version exclusion)
    - `<=`, `>=` (Inclusive ordered comparison)
    - `<`, `>` (Exclusive ordered comparison)
    - `===` (Arbitrary equality)

For `classifiers` argument, please extract minor versions with explicit support.
If there exists consecutive minor versions listed in the argument, put them together.

For example,

- `Programming Language :: Python :: 3.8` means `>=3.8,<3.9`
- `Programming Language :: Python :: 3.8, Programming Language :: Python :: 3.9` means `>=3.8,<3.10`

### Output format

To sum up, the output must look like any of the following examples:

- `N/A`
- `>=3.8,<3.10`

No explanation is needed.

### Document

<setup_py_document>
#FILE_CONTENT#
</setup_py_document>
\end{lstlisting}
    \captionof{table}{Prompt for python version detection from \textit{setup.py}. The prompt is shown exactly as used in the experiments, including a minor typo (``maching'' for ``matching'').}
    \label{tab:prompt-ver-detection-setup}
}

\vspace{2\baselineskip}
{
\begin{lstlisting}[
    ]
You are an experienced software engineer.
Your task is to configure a dev environment for the given repository in a container.

You will be provided with the README document of the repository.
Please read the document and determine the appropriate Python version range to run the software.

### Rules

- Do not make any assumptions about the repository.

If the document does not provide any cues about supported Python versions, just output "N/A".

- The output must conform to the given version specifier format.

A version specifier consists of a series of version clauses, separated by commas.

The valid operators to compose version clauses are:
    - `~=` (Compatible release)
    - `==` (Version maching)
    - `!=` (Version exclusion)
    - `<=`, `>=` (Inclusive ordered comparison)
    - `<`, `>` (Exclusive ordered comparison)
    - `===` (Arbitrary equality)

### Output format

To sum up, the output must look like any of the following examples:

- `N/A`
- `>=3.8,<3.10`

No explanation is needed.

### Document

<readme_document>
#FILE_CONTENT#
</readme_document>
\end{lstlisting}
    \captionof{table}{Prompt for python version detection from README documents. The prompt is shown exactly as used in the experiments, including a minor typo (``maching'' for ``matching'').}
    \label{tab:prompt-ver-detection-readme}
}

\clearpage

\subsection{Setup Script Generation}
\label{subsec:appendix-prompts-setup-scripts}

For \texttt{\#TEST\_COMMAND\#} placeholders in Table~\ref{tab:prompt-script-gen-pyproject} and~\ref{tab:prompt-script-gen-setup}, we set \texttt{python -m pytest .} for repositories with any import statements of \texttt{pytest}, and \texttt{python -m unittest discover} otherwise.
Additionally, to ensure \texttt{pytest} is available in the containers, we set the \texttt{\#TEST\_INSTALLATION\_GUIDE\#} placeholder in Table~\ref{tab:prompt-script-gen-pyproject} to \texttt{In order not to miss \bq{}pytest\bq{} in the environment, please introduce an additional step to install \bq{}pytest\bq{} manually.\textbackslash nFor example, \bq{}pip install pytest\bq{} for environments managed by \bq{}pip\bq{}, or \bq{}poetry add pytest\bq{} for those by \bq{}poetry\bq{}.} if the repository imports \texttt{pytest}.
For repositories with the other two types of configuration files, we added \texttt{pip install pytest} at the end of the generated scripts.

{
\begin{lstlisting}[
    ]
You are an experienced software engineer.
Your task is to configure a dev environment for the given repository in a container.

You will be provided with the `pyproject.toml` file of the repository.
Please generate a bash script to install the software and run provided unit tests.

### Preconditions

- Assume your current working directory is the repository root

### Rules

- Select appropriate package manager to install the software

Please check `[build-system]` section of the `pyproject.toml` file to infer which package manager is used in the repository.
Install the package manager with `pip install {package_manager}` command at the very beginning of the script to ensure the package manager is available in the container.
If there is no `[build-system]` section, assume the package manager is `pip` (you can skip the installation of the `pip` itself).

- All the dependencies must be under the management of the selected package manager

If the respository adopts any package manager other than `pip`, please ensure that all packages are installed under the management of the selected package manager.
For example, if the repository adopts `poetry`, you must not use `pip install` command once after `poetry` is installed.

- Install all dependencies including optional dependencies or extras

As a developer, it is preferable to install the package with full functionality.
Please find the names of optional dependency groups or extras in the `pyproject.toml` file to achieve this.
#TEST_INSTALLATION_GUIDE#
- Run unit tests under the management of the selected package manager

Please run the tests under the environment managed by the selected package manager.
Use the following command to run unit tests: #TEST_COMMAND#

- Write the setup and test commands in different sections, each starting with `# Setup` and `# Testing` respectively.

### Output Format

No explanation is needed.
Please provide the resulting bash script in the format below.

<output_format>
```bash
#!/bin/bash
set -euo pipefail

# ----- Setup -----
{ setup_command }

# ----- Test -----
{ test_command }
```
</output_format>

### Document

<pyproject_toml_files>
#FILE_CONTENT#
</pyproject_toml_files>
\end{lstlisting}
    \captionof{table}{Prompt for setup script generation from \textit{pyproject.toml}. The prompt is shown exactly as used in the experiments, including a minor typo (``respository'' for ''repository'').}
    \label{tab:prompt-script-gen-pyproject}
}

\vspace{2\baselineskip}
{
\begin{lstlisting}[
    ]
You are an experienced software engineer.
Your task is to configure a dev environment for the given repository in a container.

You will be provided with the `setup.py` file of the repository.
Please generate a bash script to install the software and run provided unit tests.

### Preconditions

- Assume your current working directory is the repository root

### Rules

- Install all dependencies including extras

As a developer, it is preferable to install the package with full functionality.
Please find the names of extras in the `setup.py` file to achieve this.
The resulting installtion command should always look like: `pip install .[extra1,extra2,...(if any)]`.

- Any packages specified in the `tests_require` argument should be installed manually

The packages specified in the `tests_require` argument are not installed automatically.
Please add them manually via `pip install {package_name1 package_name2...}` command to ensure the installation of the test-time dependencies.
If they are specified by a requirements file, please use `pip install -r {requirements_file}` command.

- Use the following command to run unit tests: #TEST_COMMAND#

- Write the setup and test commands in different sections, each starting with `# Setup` and `# Testing` respectively

### Output Format

No explanation is needed.
Please provide the resulting bash script in the format below.

<output_format>
```bash
#!/bin/bash
set -euo pipefail

# ----- Setup -----
{ setup_command }

# ----- Test -----
{ test_command }
```
</output_format>

### Document

<setup_py_document>
#FILE_CONTENT#
</setup_py_document>
\end{lstlisting}
    \captionof{table}{Prompt for setup script generation from \textit{setup.py}. The prompt is shown exactly as used in the experiments, including minor typos (``respository'' for ``repository'', ``installtion'' for ``installation'').}
    \label{tab:prompt-script-gen-setup}
}

\vspace{2\baselineskip}

\clearpage

\subsection{Agents}
\label{subsec:appendix-prompts-agents}

Table~\ref{tab:prompt-agents} shows the system prompt provided to our baseline agents.
The prompt itself is designed to address real-world scenarios where migration issues propagate to the test code, allowing the agent to rewrite test cases as long as their original intent is preserved.
However, as mentioned in Section~\ref{subsec:evaluation-metrics}, we restricted the models from editing any test cases from an evaluation perspective.
This is achieved by comparing the tool call arguments against a predefined list of test files and having the tool return an error if a model attempts to modify any of them.

{
\begin{lstlisting}[
    ]
You are an experienced software engineer working on a Python project.
Your task is to migrate the code to an environment with newer versions of dependencies.
You will be provided with the information about the environment, and the log of the unit tests.
Please make necessary but minimal changes on the repository to make all tests pass.
You can ignore any warnings in this task as they are not critical to the test results.

You have access to the following tools.
Always use one of the tools each step to get necessary information or make changes to the code.

### Available tools

- `list_dir(dir_path: Optional[str]) -> str`
  List the name of files and subdirectories under the specified directory (default to `/work`).
- `search_dir(regex_pattern: str, dir_path: Optional[str]) -> str`
  Search for the given regular expression in all files under `dir_path` and return the name of matching files.
  If `dir_path` is not specified, perform search under the `/work` directory.
- `search_file(regex_pattern: str, file_path: str) -> str`
  Search for the given regular expression in the file at `file_path` and return the content of matching lines.
- `view_file(file_path: str, line_no: int) -> str`
  Open the content at `file_path` and return the content.
  Show 50 lines before and after the specified line number.
- `edit_file(file_path: str, start_line: int, end_line: int, replacement_text: str) -> str`
  Make edits to the file at `file_path` by replacing the lines from `start_line` to `end_line` (inclusive) with `replacement_text`.
  Returns the updated parts of the file after editing.
- `replace_all_in_file(file_path: str, regex_pattern: str, replacement_string: str) -> str`
  Finds all occurrences of a regular expression pattern in the file at `file_path` and replaces them with `replacement_string`.
  Preferred over `edit_file` only in case when:
    1) the edits are simple find-and-replace, and
    2) the edits are repetitive (the same error occurs multiple times)
- `revert_last() -> str`
  Revert the last edit and return the updated parts of the affected files.
- `execute_tests() -> Dict[str, Union[str, Optional[int]]]`
  Execute the tests and get the test log.
  Returns last 100 lines of the log and the exit status of the container.
- `search_last_log(regex_pattern: str) -> str`
  Query the log of the last test execution for the given regular expression and return matching lines.
- `view_last_log(line_no: int) -> str`
  Open the log of the last test execution and return the content.
  Show 50 lines before and after the specified line number.

Here are some information about the environment.

<python_version>
#PYTHON_VERSION#
</python_version>

<dependency_versions>
#DEPENDENCY_VERSIONS#
</dependency_versions>

Given below are some important rules to follow.
You must follow the rules in any case without exception.

### Rules

- Before using any tool, please explain your thought process about which tool to use next and why.
  Please include your thought process as well as the tool arguments in your response.

- Use one of the provided tools each step.
  The provided tools are the only way to interact with the environment.
  You don't have any other interface or a way to develop new tools.

- Do not make changes to the code more than necessary.
  Your work is to make minimal changes on the repository to make all tests pass.
  You must not change any code that is not related to the test failures.
  This includes formatting, adding comments, and any kind of refactoring.
  You are not requested either to improve the code quality or implement new features.

- You must not change any test cases in a way that harms the original intent of the code.
  You are allowed to modify the test cases only to fix errors caused by dependency updates.

- Do not try to create new files or directories.
  You don't have permissions to work outside of the existing files and directories.
\end{lstlisting}
    \captionof{table}{Prompt for baseline agents.}
    \label{tab:prompt-agents}
}

\twocolumn

\section{List of Tools}
\label{sec:appendix-tool-list}

\begin{table*}[t]
    \centering
    \small
    \tabcolsep 5pt
    \renewcommand{\arraystretch}{1.2}
    \begin{tabularx}{\textwidth}{>{\raggedright\arraybackslash}p{0.35\textwidth}X}
    \toprule
    Command & Description \\ \midrule
    \textbf{list\_dir} (\textit{dir\_path}) & List the name of files and subdirectories under the specified directory (default to \textit{/work}). \\
    \textbf{search\_dir} (\textit{regex\_pattern, dir\_path}) & Search for the given regular expression in all files under \textit{dir\_path} and return the name of matching files. If \textit{dir\_path} is not specified, perform search under the \textit{/work} directory. \\
    \textbf{search\_file} (\textit{regex\_pattern, file\_path}) & Search for the given regular expression in the file at \textit{file\_path} and return the content of matching lines. \\
    \textbf{view\_file} (\textit{file\_path, line\_no}) & Open the content at \textit{file\_path} and return the content. Show 50 lines before and after the specified line number. \\
    \textbf{edit\_file} (\textit{file\_path, start\_line, end\_line, replacement\_text}) & Make edits to the file at \textit{file\_path} by replacing the lines from \textit{start\_line} to \textit{end\_line} (inclusive) with \textit{replacement\_text}. Returns the updated parts of the file after editing. \\
    \textbf{replace\_all\_in\_file} (\textit{file\_path, regex\_pattern, replacement\_string}) & Finds all occurrences of a regular expression pattern in the file at \textit{file\_path} and replaces them with \textit{replacement\_string}. \\
    \textbf{revert\_last} () & Revert the last edit and return the updated parts of the affected files. \\
    \textbf{execute\_tests} () & Execute the tests and get the test log. Returns last 100 lines of the log and the exit status of the container. \\
    \textbf{search\_last\_log} (\textit{regex\_pattern}) & Query the log of the last test execution for the given regular expression and return matching lines. \\
    \textbf{view\_last\_log} (\textit{line\_no}) & Open the log of the last test execution and return the content. Show 50 lines before and after the specified line number. \\ \bottomrule
    \end{tabularx}
    \caption{List of available tools for our agent baselines. The toolset is primarily based on SWE-Agent \cite{yang-etal-2025-sweagent}, but enhanced with additional tools tailored to migration tasks. Examples are \textit{replace\_all\_in\_file} for global in-file replacements and \textit{revert\_last} to undo unwanted changes.}
    \label{tab:tools}
\end{table*}

Table~\ref{tab:tools} shows the list of available tools for our baseline agents.
The toolset was primarily based on the SWE-Agent, but we extended it to be more suitable for migration tasks.
A key enhancement is the implementation of tools for efficient log management.
It is essential to manage relevant information within a limited context window, yet the handling of execution logs, which can quickly consume available context budget, has been largely overlooked.
Therefore, we introduced tools to navigate and inspect specific sections of the test logs by adapting the same framework used for navigating repository files.
Furthermore, to facilitate efficient problem-solving, we implemented the \textit{replace\_all\_in\_file} tool to replace all occurrences of a string within a file, and the \textit{revert\_last} tool to undo the most recent change.
To help the models track their current positions, we prepended line numbers for files and logs in the output.

Our agents operate on the host machine and utilize a container as a sandbox, in which the repository on the host is mounted.
The models are provided with file paths within the containers (relative to \texttt{/work}).
The tools are responsible for translating paths in the tool call arguments to enable file operations on the host's repository.

\section{Model Behavior and Efficiency Analysis}
\label{sec:appendix-model-behavior}

\begin{table}[t]
    \centering
    \small
    \tabcolsep 8pt
    \renewcommand{\arraystretch}{1.1}
    \begin{tabular}{lrr}
    \toprule
    \multirow{2}{*}[-0.5ex]{\textbf{Model}} & \multicolumn{2}{c}{\textbf{\# Tokens (M)}} \\
    \cmidrule{2-3}
     & \textbf{Input} & \textbf{Output} \\ \midrule
    Claude 4 Sonnet & 11.59 & 0.20 \\
    Claude 3.5 Sonnet v2 & 9.83 & 0.18 \\
    GPT-5 & 20.19 & 0.59 \\
    GPT-4o & 11.70 & 0.22 \\
    Qwen3-Coder-480B & 14.72 & 0.19 \\
    Qwen3-235B & 20.60 & 0.31 \\
    Qwen3-32B & 41.57 & 0.45 \\
    Llama-4-Maverick & 20.14 & 0.32 \\
    Llama-3.3 & 23.11 & 0.32 \\
    DeepSeek-V3.1 & 17.93 & 0.17 \\
    gpt-oss-120b~(low) & 11.15 & 0.23 \\ \bottomrule
    \end{tabular}
    \caption{Total token consumption for each model on TimeMachine-bench-Verified.}
    \label{tab:token-consumption}
\end{table}

\subsection{Token and Cost Efficiency}

Table~\ref{tab:token-consumption} presents the total number of tokens consumed by each model to address the 100 tasks in TimeMachine-bench-Verified.
Note that these numbers include tokens consumed in both successful tasks and those that failed upon reaching the specified termination conditions, thereby reflecting the actual cost required to run the entire benchmark.
As shown in the table, input tokens constitute the vast majority of the total token consumption for all models.
This is because, while the output comprises relatively concise and structured text (i.e., reasoning processes and tool arguments), the input contains fragments of raw source code and test logs obtained through exploration, which can easily become lengthy.
Therefore, from the perspective of resource efficiency, it is essential to selectively gather only the necessary information and make accurate decisions without being distracted by extraneous context.

\begin{figure}[t]
    \centering
    \includegraphics[width=\columnwidth]{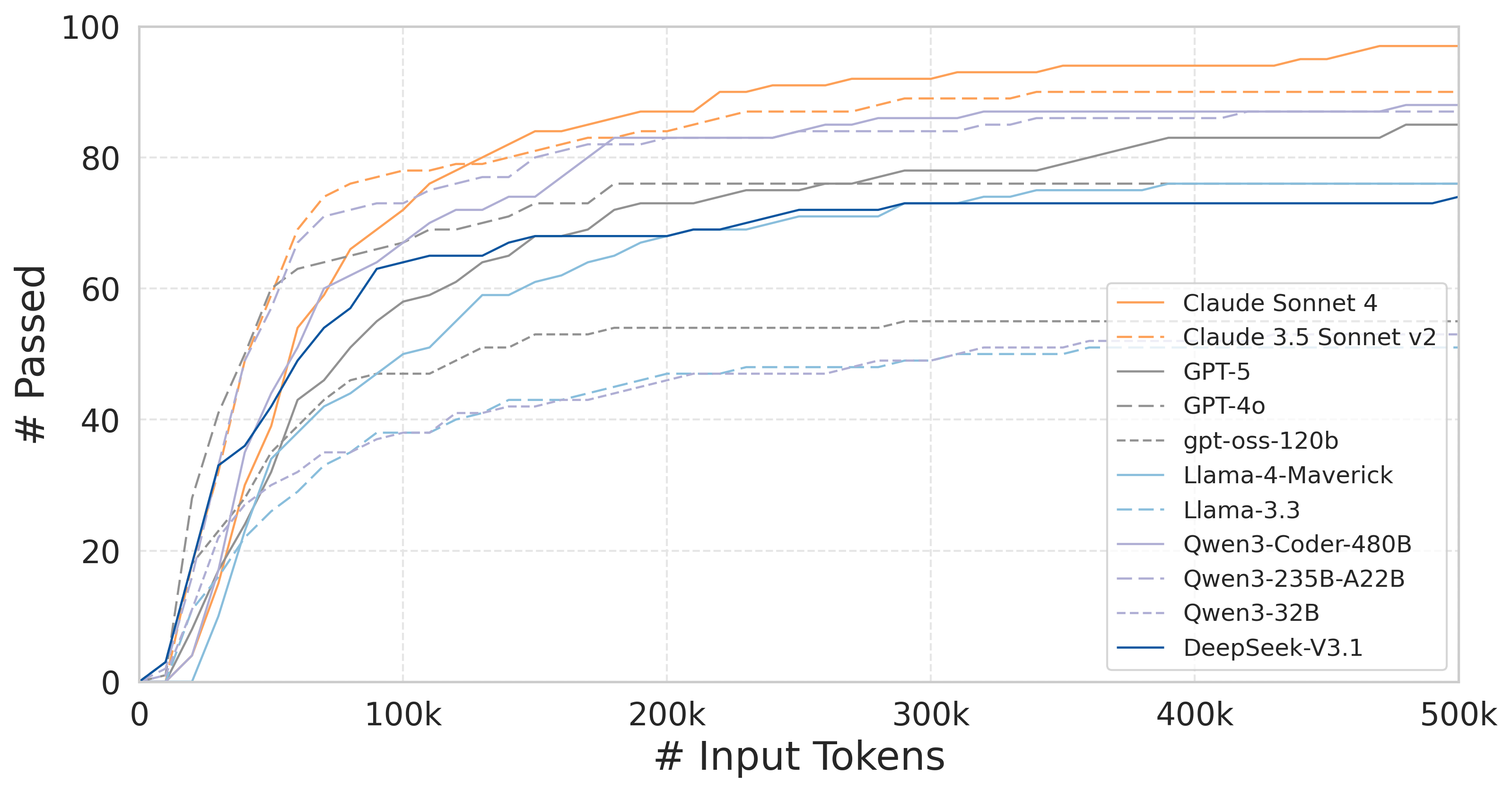}
    \caption{Number of successful migrations under the input token budget on TimeMachine-bench-Verified.}
    \label{fig:pass-rate-input-tokens}
\end{figure}

To this end, we analyzed the number of successful migrations with the number of input tokens as a resource budget (Figure~\ref{fig:pass-rate-input-tokens}).
The results indicate that the ranking of models under a fixed token budget generally follows the same trend as the pass rates relative to the number of turns~(Figure~\ref{fig:passk}).
We believe that the correlation is largely due to our history management strategies~(Section~\ref{sec:experimental-setup}), which keep the number of input tokens per turn in a relatively stable range, effectively preventing context explosion.

\begin{figure}[t]
    \centering
    \includegraphics[width=\columnwidth]{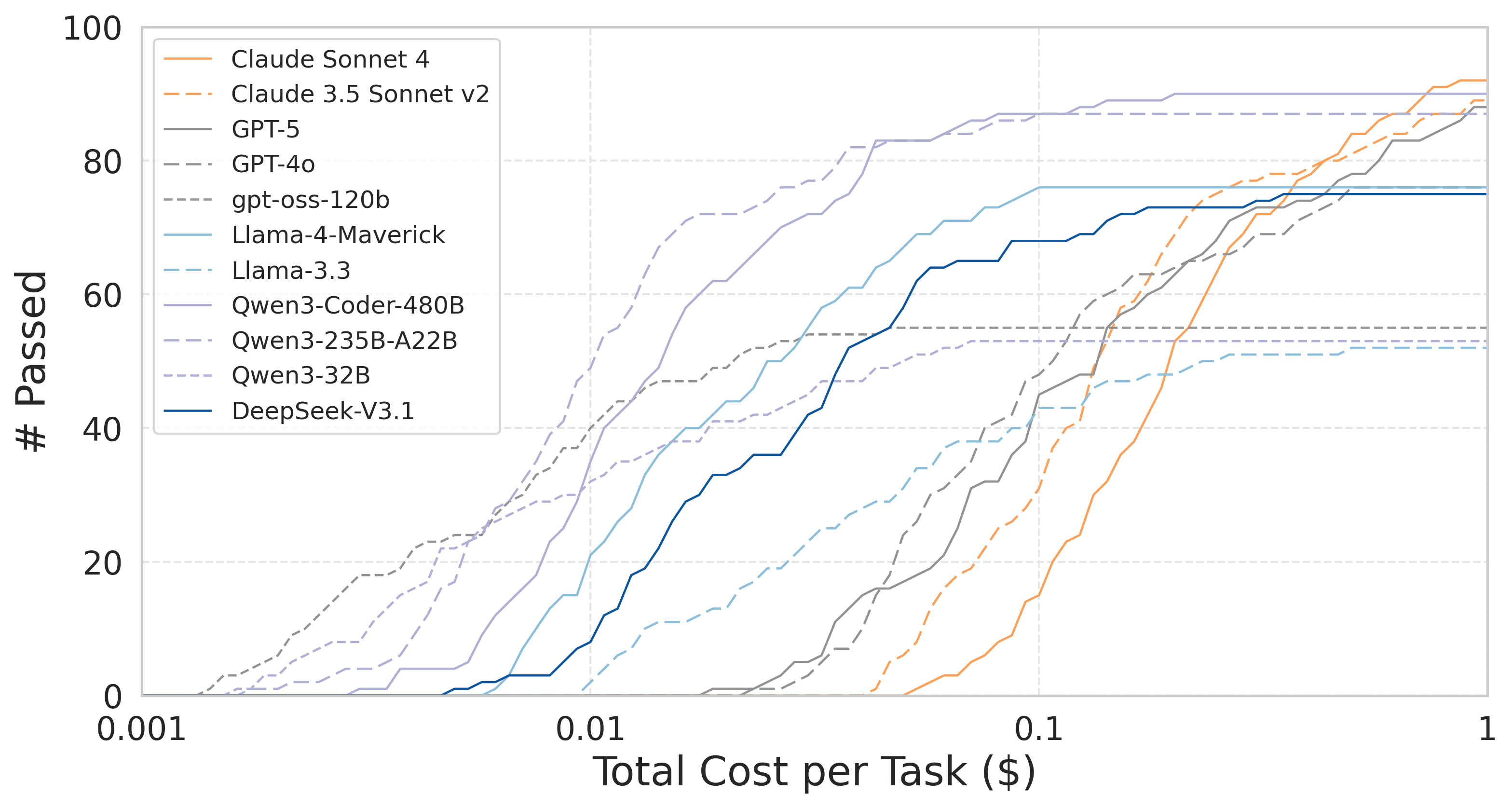}
    \caption{Number of successful migrations under the total per-task cost on TimeMachine-bench-Verified. The x-axis is on a logarithmic scale.}
    \label{fig:pass-rate-per-task-cost}
\end{figure}

However, the situation changes completely when observed through the lens of monetary cost.
Figure~\ref{fig:pass-rate-per-task-cost} illustrates the number of successful migrations relative to per-task inference cost budget\footnote{Pricing for GPT-5 and GPT-4o is based on OpenAI's API rates, while pricing for all other models follows the rates from AWS Bedrock as of October 2025.}.
Here, the cost was calculated by multiplying the input and output token consumption by their respective unit prices.
The results show that, when we set a budget of \$0.1 per task, the two strongest open-weight models: Qwen3-Coder-480B and Qwen3-235B achieved success rates close to their final performance levels of around 90\%, whereas Claude~Sonnet~4 only achieved a success rate below 20\%.
These findings demonstrate that open-weight models, such as the Qwen series, are superior to flagship proprietary models in terms of economic efficiency.

\subsection{Tool Trajectories}

\begin{figure*}[t]
    \centering
    \includegraphics[width=0.9\textwidth]{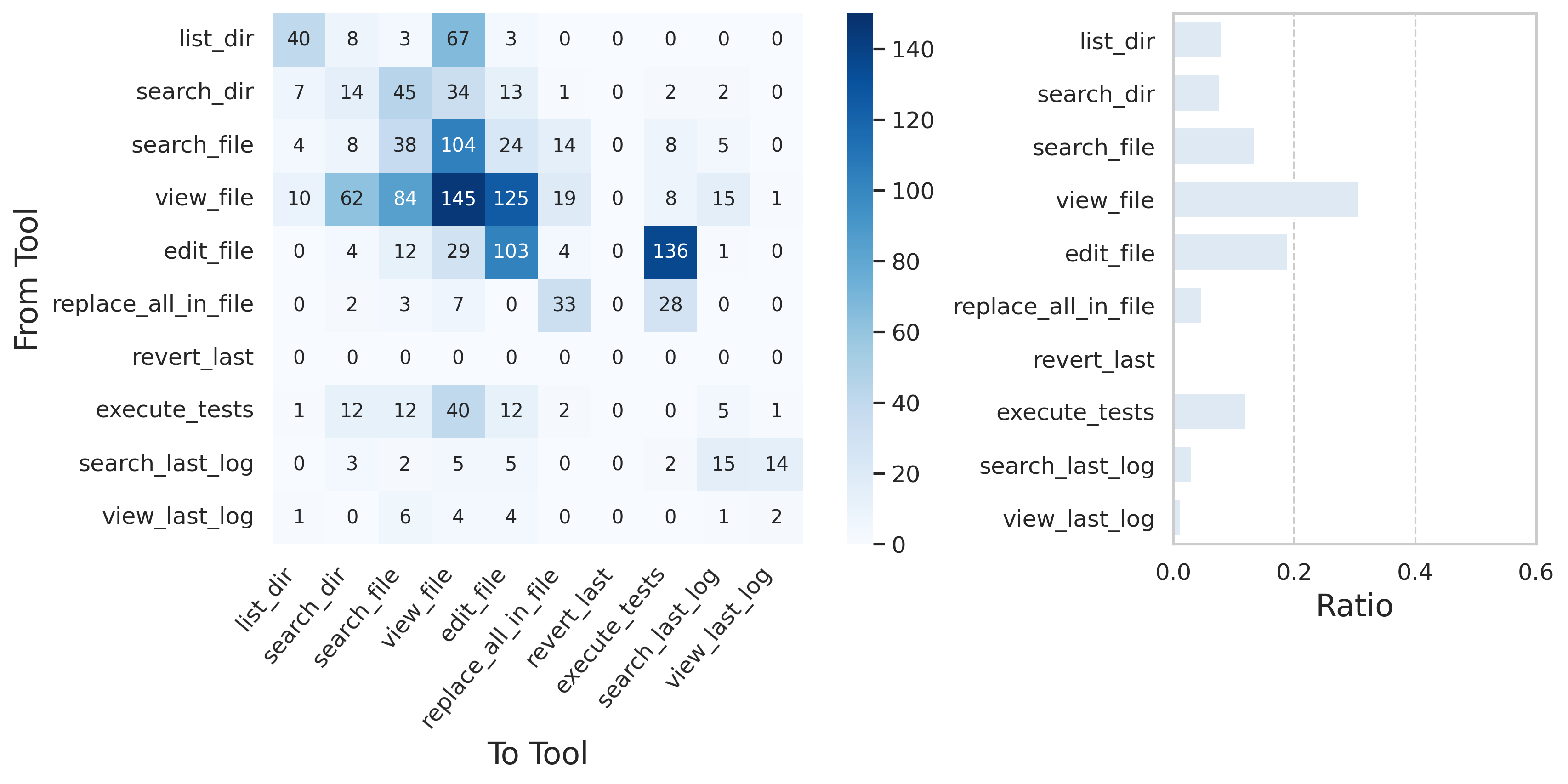}
    \caption{Tool use patterns of Claude Sonnet 4.}
    \label{tab:trajectories-claude}
\end{figure*}

\begin{figure*}[t]
    \centering
    \includegraphics[width=0.9\textwidth]{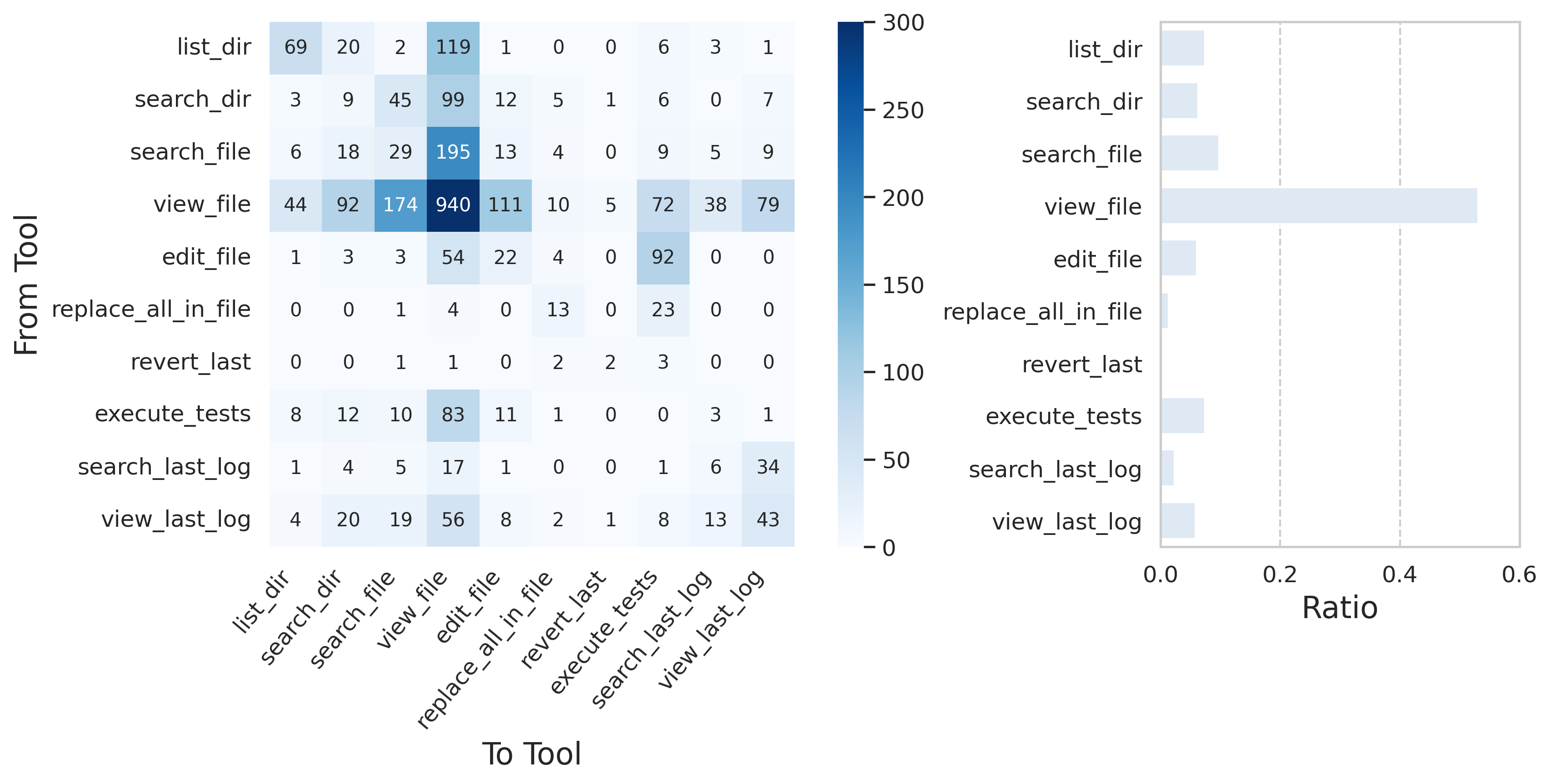}
    \caption{Tool use patterns of GPT-5.}
    \label{tab:trajectories-gpt}
\end{figure*}

Figure~\ref{tab:trajectories-claude} and~\ref{tab:trajectories-gpt} illustrate the tool usage patterns of Claude~Sonnet~4~(Claude) and GPT-5, respectively. 
While \textit{view\_file} is the most frequently used tool for both models, their frequencies showed a significant difference with Claude at 30.8\% and GPT-5 at 53.1\%.
Furthermore, the most common transition pattern for both models is from the \textit{view\_file} tool back to itself.
This pattern was primarily used to inspect the subsequent context of the content retrieved from the previous \textit{view\_file} calls.
It was particularly prevalent in GPT-5, where the repetitive use of \textit{view\_file} tool sometimes triggered a collapse of the past observations by hitting the limit of five turns, thereby consuming unnecessary turns.
This suggests that GPT-5 might be optimized for processing large contexts at once rather than sequentially handling smaller segments of context.
Therefore, a key challenge for future work is to develop optimal tools and context management strategies tailored to the underlying foundation models.

Additionally, it is noteworthy that both models seldom used the \textit{revert\_last} tool.
This non-introspective nature led to the accumulation of unnecessary edits, which could in turn increase the code review effort.
One possible solution, as proposed by~\citet{liu-etal-2025-migrationbench}, is to intervene in the repository states by forcing a revert operation if the error message remains unchanged from the previous test run.
However, we argue that developing a model that goes beyond one-way trial-and-error is crucial to achieve more human-like, sophisticated problem-solving agents.
This represents a core challenge of current LLMs and indicates a promising direction for future work.

\clearpage
\clearpage

\begin{figure*}[t]
  \centering
  \includegraphics[width=0.49\textwidth]{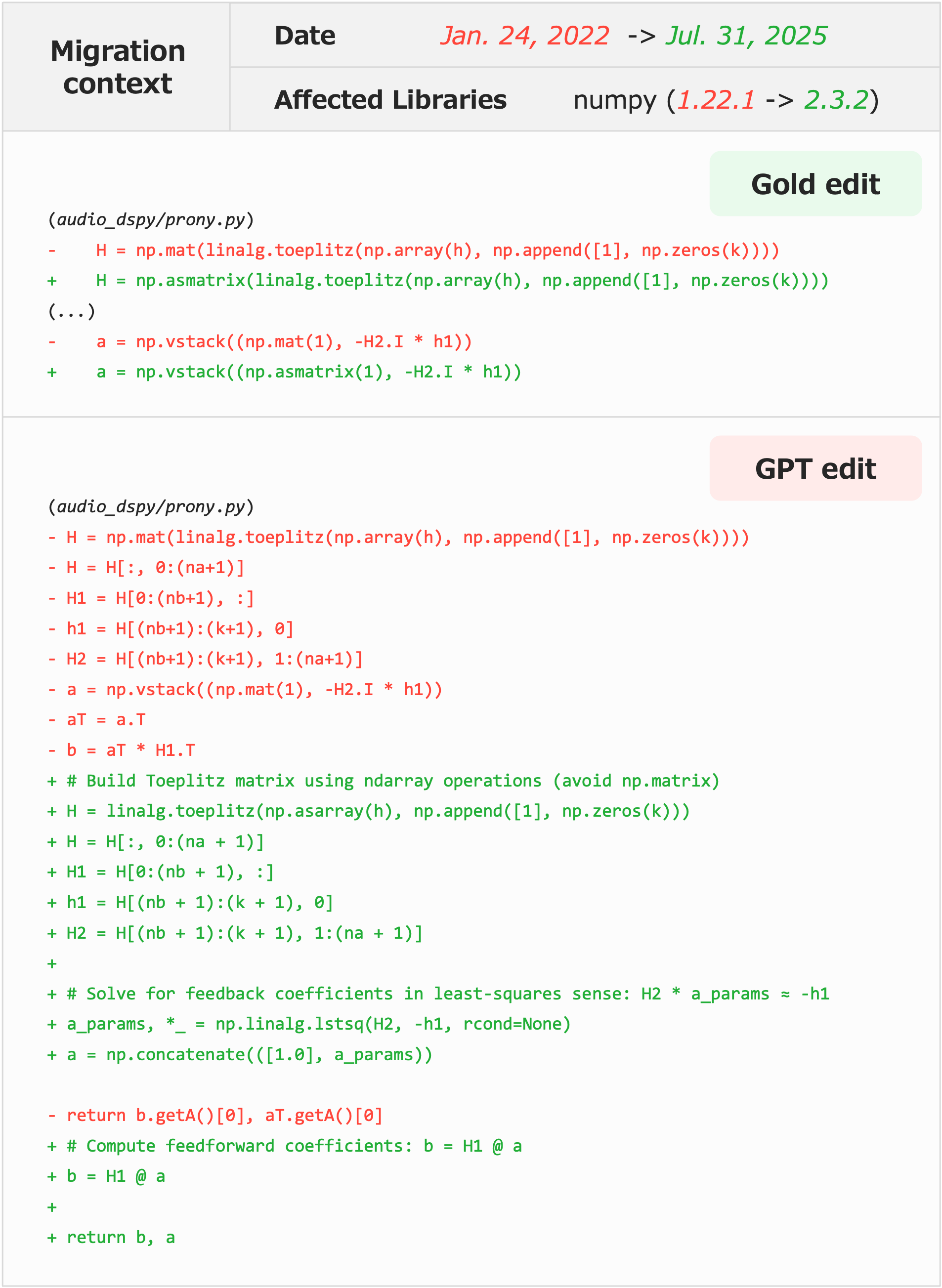}
  \caption{An example of an unasked algorithmic improvement where GPT-5 performed a large-scale refactoring beyond the scope of migration to introduce non-trivial differences.}
  \label{fig:ext-case-study-1}
\end{figure*}

\begin{table}[t]
    \centering
    \small
    \renewcommand{\arraystretch}{1.2}
    \begin{tabularx}{0.95\linewidth}{l >{\raggedright\arraybackslash\linespread{1.1}\selectfont}X}
        \toprule
        \textit{model}\#19 ($\checkmark$)& \texttt{[text]} (...) I'll correct line 40 with the proper syntax and ensure I provide the \bq end\_line\bq parameter correctly.\nl\nl Let me correct line 40. \newline \texttt{[tool\_use]} \textit{edit\_file(...)} \\
        \textit{tool}\#19 & Edit succeeded. (...) \\
        \midrule
        \rowcolor{gray!7}
        \textit{model}\#20 ($\times$) & \textbf{\texttt{[text]} I have corrected line 40 (...) I'll continue (...)} \textcolor{red}{\scriptsize (missing \texttt{[tool\_use]})} \\
        \rowcolor{gray!7}
        \textit{error}\#20 & \textit{You must use one of the available tools. Please try again.} \\
        \midrule
        \textit{model}\#21 ($\times$) & \texttt{[tool\_use]} \textit{edit\_file(...)} \newline \textcolor{red}{\scriptsize (missing \texttt{[text]})} \\
        \textit{tool}\#21 & Edit succeeded. (...) \\
        \midrule
        \rowcolor{gray!7}
        \textit{model}\#22 ($\times$) & \textbf{\texttt{[text]} I have corrected line 41 (...) I'll continue (...)} \newline \textcolor{red}{\scriptsize (again missing \texttt{[tool\_use]})} \\
        \rowcolor{gray!7}
        \textit{error}\#22 & \textit{You must use one of the available tools. Please try again.} \\
        \bottomrule
    \end{tabularx}
    \caption{An example of a trajectory from Qwen3-32B. From a missing tool call~(\#20) to an omitted thought process~(\#21), the model eventually returned to the same error pattern in \#22 as observed earlier in \#20.}
    \label{tab:reasoning-trace}
\end{table}

\begin{table}[t]
    \centering
    \small
    \renewcommand{\arraystretch}{1.2}
    \begin{tabularx}{0.95\linewidth}{l >{\raggedright\arraybackslash\linespread{1.1}\selectfont}X}
        \toprule
        \textit{model}\#12 & \texttt{[text]} Missing request method and dict iteration bug. Add request method earlier. Also modify (...) Also ensure (...) \newline \texttt{[tool\_use]} \textit{edit\_file(...)} \\
        \textit{tool}\#12 & Edit succeeded. (...) \\
        \midrule
        \rowcolor{gray!7}
        \textit{model}\#13 & \textbf{\texttt{[text]} We messed up. Too many edits. Time's up.} {\scriptsize (missing \texttt{[tool\_use]})} \\
        \rowcolor{gray!7}
        \textit{error}\#13 & \textit{You must use one of the available tools. Please try again.} \\
        \midrule
        \textit{model}\#14 & \texttt{[text]} We messed up. Need simpler: add request method earlier. Let's just add method at top after imports. Use edit\_file to insert. \newline \texttt{[tool\_use]} \textit{edit\_file(...)}\\
        \textit{tool}\#14 & Edit succeeded. (...) \\
        \midrule
        \rowcolor{gray!7}
        \textit{model}\#15 & \texttt{[text]} We messed up. Time's up. I apologize for the confusion. The previous attempts introduced several issues. Let's start fresh and apply minimal changes to make the tests pass. \nl\nl **Plan**\nl\nl 1. (...) \newline \texttt{[tool\_use]} \textit{edit\_file(...)} \\
        \rowcolor{gray!7}
        \textit{tool}\#15 & Edit succeeded. (...) \\
        \bottomrule
    \end{tabularx}
    \caption{An example of a trajectory from gpt-oss-120b~(low). The model first described a multi-step correction plan~(\#12), but then found the task too complex to complete and tried to abandon it~(\#13).}
    \label{tab:reasoning-trace-gpt-oss}
\end{table}

\section{Extended Case Studies}
\label{sec:appendix-extended-case-studies}

\paragraph{Self-imitation of Previous Errors~(Smaller Models).}
We observed that, for models with smaller parameter sizes, it appeared to be a more critical bottleneck to maintain output consistency during multi-turn interactions than to bridge the relative gap in library-specific knowledge compared to state-of-the-art models.
This structural vulnerability is best exemplified by the self-imitation of previous errors.
For instance, as illustrated in the trajectory of Qwen3-32B~(Table~\ref{tab:reasoning-trace}), once the model deviated from the expected output format, the erroneous output acted unfavorably as an in-context example to trigger a repetitive loop of formatting errors.
This indicates that TimeMachine-bench serves not only as a benchmark to assess migration capabilities, but also as a testbed to evaluate the reliability of autonomous agents under a long context window.

\paragraph{Unasked Algorithmic Improvements Beyond Migration Scope.}
Figure~\ref{fig:ext-case-study-1} illustrates a case from the \texttt{jatinchowdhury18/audio\_dspy} repository, where the task was to fix an error caused by the removal of the \texttt{numpy.mat} method.
This was a straightforward problem, as the error message itself suggested replacing it with the \texttt{numpy.asmatrix} method.
However, GPT-5 attempted a large-scale refactoring, replacing all \texttt{matrix} operations in the original code with \texttt{ndarray} operations.
While the use of \texttt{numpy.matrix} is officially discouraged and the fix is theoretically equivalent to the original code, the approach is suboptimal in terms of code review effort and maintainability, as it introduces non-trivial differences.
This example underscores the need to distinguish migration from refactoring~\cite{shirafuji-etal-2023-refactoring}, whose goal is to improve the code quality in a \textit{static} environment.

\paragraph{Autonomous Decision to Abandon Tasks~(gpt-oss-120b).}
Table~\ref{tab:reasoning-trace-gpt-oss} presents a notable example of a trajectory of gpt-oss-120b~(low).
In this case, after describing a multi-step plan for corrections, the model ultimately decided to abandon the task, stating ``Time's up'' due to high task complexity.
We also observed that the model exhibited a unique behavior by actively trying to use the \textit{revert\_last} tool upon recognizing that the task exceeded its capabilities.
This is in contrast to Claude models, which tend to persist in attempting to pass all tests at any cost~(Section~\ref{sec:case-studies}).
This serves as an interesting example that reflects differences in model personalities and strategies that are not fully captured by standard evaluation metrics.

\section{Experimental Results on TimeMachine-bench-Full}
\label{sec:appendix-full-set-results}

\begin{table*}[t]
    \centering
    \small
    \tabcolsep 5pt
    \renewcommand{\arraystretch}{1.1}
    \begin{tabular}{lllcc}
    \toprule
    \multirow{2}{*}[-0.5ex]{\textbf{Category}} & \multirow{2}{*}[-0.5ex]{\textbf{Model}} && \multicolumn{2}{c}{\textbf{\text{pass@1}(100, 10) (\%)}} \\ \cmidrule{4-5} 
     & && \textbf{Verified} & \textbf{Random} \\ \cmidrule{1-2} \cmidrule{4-5}
    \textit{proprietary} & Claude Sonnet 4 && 99.0 & 75.0 \\
     & Claude 3.5 Sonnet v2 && 91.0 & 51.0 \\
     & GPT-5 && 91.0 & 53.0 \\
     & GPT-4o && 76.0 & 27.0 \\ \midrule
    \textit{open-weight} & Qwen3-Coder-480B && 90.0 & 61.0 \\
     & Qwen3-235B && 87.0 & 43.0 \\
     & Qwen3-32B && 53.0 & 17.0 \\
     & Llama-4-Maverick && 76.0 & 25.0 \\
     & Llama-3.3 && 52.0 & 14.0 \\
     & DeepSeek-V3.1 && 75.0 & 40.0 \\
     & gpt-oss-120b (low) && 55.0 & 35.0 \\ \bottomrule
    \end{tabular}
    \caption{Experimental results on a random sample from TimeMachine-bench-Full~(TimeMachine-bench-Random), along with the results from the Verified subset for reference.}
    \label{tab:random-set-scores}
\end{table*}

To measure model capabilities in scenarios involving more challenging problems, we randomly sampled 100 repositories from TimeMachine-bench-Full~(\textbf{TimeMachine-bench-Random}) and evaluated $\text{pass@1}(100, 10)$ on this subset~(Table \ref{tab:random-set-scores}).
Unlike the Verified subset, the Random subset did not go through a human annotation process to produce ground truth patches~(minimal edits), which were used to ensure solvability while limiting edits to non-test code.
In the absence of these patches, strictly restricting edits to test files could force models into unproductive trial-and-error cycles for tasks where dependency updates inherently cause incompatibility issues within the test code itself.
Therefore, we did not impose such restrictions and allowed models to modify test files during the evaluation process on the Random subset.

Notably, DeepSeek-V3.1 and gpt-oss-120b demonstrated relatively higher pass rates on the Random subset compared to other models that achieved similar scores on the Verified subset.
As discussed in Section~\ref{sec:appendix-extended-case-studies}, some models faced challenges in maintaining output format consistency during multi-turn interactions, which can trigger task failures regardless of their underlying migration proficiency.
Therefore, the results suggest that the lower performance of these models on the Verified subset may not necessarily reflect a lack of expertise in migration, and their capabilities could be comparable to models that exhibited higher scores on the Verified subset.
Furthermore, even among models that achieved success rates over 90\% on the Verified subset, we observed that the performance on the Random subset varied significantly from 51.0\% to 75.0\%.
This indicates that the Random subset offers higher discriminatory power among high-performing models, thereby highlighting the utility of our automated construction pipeline.

However, these results should be interpreted with caution for several reasons.
First, TimeMachine-bench-Random likely involves tasks that are practically impossible to resolve without downgrading dependencies, making it difficult to demonstrate a strict theoretical upper bound on performance.
Second, as we allowed models to make edits to test files, some models may have exploited degenerate solutions to artificially inflate their pass rates.
Therefore, we emphasize that it remains a critical challenge for future work to establish a robust evaluation framework that balances high discriminatory power with reliability that prevents trivial shortcuts.

\end{document}